\documentclass[showpacs,prd,nofootinbib,showkeys,twocolumn,10pt,aps,longbibliography]{revtex4-1}
\usepackage{hyperref} 
\usepackage{amsmath,amssymb}
\usepackage{graphicx}
\usepackage[caption=false]{subfig}
\usepackage{comment}
\usepackage{color}

\definecolor{darkgreen}{rgb}{0,.7,0}
\definecolor{linkblue}{rgb}{0.,0.,0.9333}

\hypersetup{
    pdffitwindow=true,      
    pdfauthor={N. N. Barnard},     
    pdfnewwindow=true,      
    bookmarksnumbered=true,
    colorlinks=true,       
    linkcolor=red,          
    citecolor=darkgreen,        
    filecolor=magenta,      
    urlcolor=cyan,           
    menucolor=blue,
    baseurl=http://www.arxiv.org/abs/,
}

\newcommand{\pt}{$p_T$}

\newcommand{\highpt}{high-\pt{}}

\newcommand{\lowpt}{low-\pt{}}

\begin{document}

\title{Bottomonia Suppression in Heavy Ion Collisions from AdS/CFT}

\author{N.\;N.\;Barnard}
\email{brnnad007@myuct.ac.za}
\affiliation{Department of Physics, University of Cape Town, Private Bag X3, Rondebosch 7701, South Africa}
\author{W.\;A.\;Horowitz}
\email{wa.horowitz@uct.ac.za}
\affiliation{Department of Physics, University of Cape Town, Private Bag X3, Rondebosch 7701, South Africa}

\date{\today}

\begin{abstract}
We compute for the first time the suppression of bottomonia in a strongly coupled QGP and compare the results to those from a weakly coupled QGP.  Using imaginary time techniques we numerically determine the real and imaginary parts of the ground state binding energy of the bottomonia in one potential computed from AdS/CFT and another computed from pQCD.  We then use these binding energies in a suppression model to determine the $\Upsilon$(1S) nuclear modification factor in $\sqrt{s}_{NN}=2.76$ TeV Pb+Pb collisions.  AdS/CFT significantly overpredicts the suppression compared to data, although the predictive power of our calculation is limited by its significant sensitivity to the exact details of the suppression model.
\end{abstract}

\maketitle

\section{\label{sec:level1}Introduction}
There is abundant evidence that high multiplicity collisions at the Relativistic Heavy Ion Collider (RHIC) and the Large Hadron Collider (LHC) yield qualitatively new physics never before seen at previous colliding energies \cite{Adams:2005dq,Adcox:2004mh,Aad:2010bu,Khachatryan:2010gv,Aad:2012gla,Abelev:2012ola,Chatrchyan:2012lxa}.  
A natural explanation for these novel observations is that in these high multiplicity collisions a new phase of colored matter is created, known as the quark-gluon plasma (QGP), in which the relevant degrees of freedom for the strong nuclear force are no longer hadrons \cite{Gyulassy:2004zy}.  
Currently, divergent ideas for the relevant dynamics of the QGP medium in high multiplicity hadronic collisions have had success in describing various experimental observables.  For example, assuming the medium is best described using the strong coupling techniques of AdS/CFT explains the success of the picture of a rapidly thermodynamizing, nearly inviscid fluid that couples strongly to high transverse momentum (\highpt{}) open partons \cite{Teaney:2003kp,Chesler:2010bi,Song:2010mg,Gale:2012rq,Bernhard:2016tnd,Alqahtani:2017tnq,Morad:2014xla,Horowitz:2015dta,Hambrock:2017sno,Brewer:2017dwd,Weller:2017tsr}.  Simultaneously, weak coupling techniques from perturbative quantum chromodynamics (pQCD) have also shown success in describing the distribution of both \lowpt{} and \highpt{} particles in high multiplicity collisions \cite{Molnar:2001ux,Lin:2001zk,Bzdak:2014dia,Koop:2015wea,He:2015hfa,Lin:2015ucn,Gyulassy:2001nm,Vitev:2002pf,Wang:2002ri,Majumder:2004pt,Dainese:2004te,Armesto:2005mz,Wicks:2005gt,Majumder:2007ae,Zhang:2009rn,Vitev:2009rd,Schenke:2009gb,Young:2011qx,Majumder:2011uk,Horowitz:2011gd,Buzzatti:2011vt,Horowitz:2012cf,Djordjevic:2013xoa}.  Even models that assume strong coupling dynamics for \lowpt{} medium modes that are weakly coupled to \highpt{} modes have been argued as successfully describing LHC jet data \cite{Liu:2006ug,Casalderrey-Solana:2014bpa,Casalderrey-Solana:2015vaa,Casalderrey-Solana:2016jvj}.  

In vacuum, quarkonia are bound states of a heavy quark and its anti-quark pair, e.g.\ the $c\bar{c}$ pair in a $J/\psi$ meson or a $b\bar{b}$ pair in the $\Upsilon$(1S) meson \cite{Olive:2016xmw}.  Embedded in a medium, the properties of quarkonia change.  Matsui and Satz \cite{Matsui:1986dk} were the first to propose that suppression of the $J/\psi$ meson spectrum should be observed in the QGP due to Debye-screening of the color charge. Heavy quarkonia may theoretically exist in conjunction with the QGP at $T>T_c$, where $T_c$ is the critical temperature required for QGP formation, due to its small binding radii relative to the screening radius, whereas lighter hadrons dissociate at $\sim T_c$. At some $T$, the screening radius eventually becomes smaller than the typical heavy quarkonia radii, leading to their dissolution \cite{Karsch:1987pv}. In addition, excited states of heavy quarkonia dissociate before the ground state \cite{Karsch:2005nk}. The suppression of the bound states of heavy quarkonia in heavy-ion collisions is hence a valuable indicator of the formation of QGP, and the comparison of the quarkonia spectra in high multiplicity collisions to that in minimum bias $p+p$ collisions where no QGP is formed is a useful probe of the QGP's properties. 

Also pioneered in \cite{Matsui:1986dk} was the use of potential models to describe the interaction of the quark and antiquark in the $q\bar{q}$ pair to calculate the suppression of quarkonia spectra in heavy-ion collisions. In these potential models, the large mass and small relative velocity of the heavy quarks justifies the use of non-relativistic quantum mechanics to describe the quarkonia: the non-relativistic Schr\"odinger equation gives a binding energy for the quarkonia given a model potential for the $q\bar{q}$ interaction.  

Further works \cite{Laine:2006ns,Beraudo:2007ky,Burnier:2007qm,Laine:2007qy,Rothkopf:2011db} have shown that in addition to a standard real Debye-screened term, the potential of heavy quarkonia at finite temperature contains an imaginary part which gives the thermal width of the state, and hence its suppression. One of the first to do so was \cite{Laine:2006ns}, which made use of perturbative methods to find the static potential of heavy quarkonia at finite temperature. They concluded that the thermal width of the state increases with temperature $T$, suggesting that at high $T$ the dissociation due to the effect described by the imaginary part of the potential occurs before color screening can even come into effect. Physical interpretations of this damping were provided by \cite{Beraudo:2007ky,Burnier:2007qm} in the form of  inelastic scatterings of hard particles in the plasma with one another and surrounding gluons, with \cite{Brambilla:2008cx} suggesting that quarkonia suppression could also be attributed to $q\bar{q}$ color singlet to octet break-up even when the Debye mass is smaller than the quarkonia binding energy.

The complex-valued potential was explored further using non-perturbative lattice QCD by \cite{Rothkopf:2011db,Laine:2007qy}, allowing for the study of strongly coupled quarkonia as well. However, lattice techniques are restricted to quarkonia at or very near rest with respect to the medium and inverting lattice observables to the quarkonia potential is a non-trivial, possibly ill-defined process \cite{Andronic:2015wma,Aarts:2016hap}; certainly another theoretical tool is required to investigate a quarkonium moving rapidly with respect to a strongly coupled medium.

 The potential for static heavy quarkonia at finite temperature in $\mathcal{N}=4$ super Yang-Mills theory was calculated via the methods of AdS/CFT by \cite{Rey:1998bq,Brandhuber:1998bs,Albacete:2008dz,Noronha:2009da,Hayata:2012rw,Fadafan:2015kma}. Liu, Rajagopal and Wiedemann (LRW)~\cite{Liu:2006nn} were the first to present a description from AdS/CFT of the consequences of velocity on the screening length of charmonium.  They found that the plasma screening length decreased with velocity and therefore could result in a significant additional source of suppression at high transverse momentum $p_T$. Since then, \cite{Finazzo:2014rca,Braga:2016oem,Patra:2015qoa,Ali-Akbari:2014vpa} have performed similar investigations. While the aforementioned are limited in their scope of application, it is interesting to note that \cite{Finazzo:2014rca} in particular concludes that the effect of velocity may not be as consequential as postulated in LRW.

On the other hand, from pQCD, \cite{Escobedo:2013tca} found a potential for weakly coupled heavy quarkonium states which is dependent on velocity and shows that the dissociation energy increases with quarkonia velocity.

We would ultimately like to investigate the consequences of these different velocity dependence pictures from AdS/CFT compared to pQCD.  In this paper we have a more modest goal: to compare the experimentally measurable consequences of pQCD vs.\ AdS/CFT pictures for computing the quarkonia potential when the quarkonia are at rest with respect to the medium.

The usual observable used to compare theoretical predictions for quarkonia to data is the nuclear modification factor, $R_{AA}(\{x_i\})$, which is the ratio of the number of quarkonia observed in an $A+A$ hadronic collision as a function of the set of variables $\{x_i\}$ to the number observed in minimum bias $p+p$ collisions, scaled by the number of $p+p$--like collisions in the $\{x_i\}$ collisions.  Should the production processes for quarkonia remain unchanged in an $A+A$ collision, then an $R_{AA}$ less than 1 indicates a suppression of quarkonia and would indicate the presence of a medium that caused the quarkonia to dissociate.  

One of the challenges of quarkonia research, however, is the significant number of unknowns that cloud the interpretation of $R_{AA}$.  For example, even in $p+p$ collisions, the production mechanism for quarkonia is not under good theoretical control \cite{Andronic:2015wma,Aarts:2016hap}; thus it is unclear how the production spectrum of quarkonia is affected quantitatively in $A+A$ collisions.  By focusing on bottomonia, whose formation time $\sim1/2m_b\ll\tau_{form}$, where $\tau_{form}$ is the formation time of the QGP medium in $A+A$ collisions, we hope to limit our theoretical uncertainty due to quarkonia formation physics.

Another complication is the possibility for regeneration.  As $\surd s$ and $T$ increase, the density of open charm in the medium increases.  Then the possibility of quarkonia to spontaneously form from these charm combining or of dissociated quarkonia reforming through the capture of these open in-medium charm quarks increases.  Since $m_b\gg T$ at RHIC and LHC and the hard production cross section for bottom is small enough, we can avoid considering regeneration in our bottomonia calculations \cite{Krouppa:2015yoa}.  

In this paper we consider the case of ground state bottomonia at rest with respect to an isotropic quark-gluon plasma. We follow the methodology outlined in Krouppa et al.\ \cite{Krouppa:2015yoa}, with a number of improvements.  Given a potential, we evolve a random wave function through imaginary time; after a sufficiently long evolution, only the ground state wave function remains. This ground state wave function then determines the ground state binding energy.  We independently confirmed these binding energies by an application of the complex Ritz variational method \cite{Kraft2013}.  Following Krouppa, we used the complex binding energies in a quarkonia suppression model to compute $R_{AA}$.  

The potential models used are presented in Section~\ref{sec:Pot}, taken from leading-order perturbative calculations in \cite{Krouppa:2015yoa} for weakly coupled quarkonia and AdS/CFT calculations in \cite{Albacete:2008dz} for strongly coupled quarkonia. These potentials are used to solve the non-relativistic, time dependent Schr\"odinger Equation (TDSE) for the ground state energy and hence binding energy of $\Upsilon$(1S), using the imaginary time techniques as outlined in Section~\ref{sec:TDSE}. The resulting binding energies as a function of temperature are provided in Section~\ref{sec:RES} and are independently confirmed using complex variational methods as laid out in Appendix~\ref{app:RITZ}. Our predictions for $R_{AA}$ as a function of $N_{\mathrm{part}}$ and $p_T$, respectively, follows in Section~\ref{sec:RAA}. We conclude our manuscript with a Discussion and Outlook in Section~\ref{sec:DISC}.  

\section{Potential Models}\label{sec:Pot}

\subsection{Weakly Coupled Quarkonia}\label{Strick}
The potential model presented here for weakly coupled quarkonia is taken from \cite{Krouppa:2015yoa} and is complex-valued as $V(r)=\Re[V(r)]+i\Im[V(r)]$. Both the real and the imaginary parts of the potential were found using leading-order perturbative calculations.  The real part of the potential was derived from the Fourier transform of the gluon propagator in the real time formalism, and the imaginary part of the potential was derived from the symmetric propagator in the imaginary time formalism; these derivations are presented in \cite{Strickland:2011aa} and \cite{Dumitru:2009fy}, respectively.

The real part of the potential is given by
\begin{eqnarray}\label{ReVStr}
	\Re[V(r)]=-\frac{\alpha}{r}(1+\mu r)e^{-\mu r}+&&\frac{2\sigma}{\mu}(1-e^{-\mu r})\nonumber\\
	&&-\sigma r e^{-\mu r}-\frac{0.8\sigma}{m_Q^2 r},
\end{eqnarray}
where $r$ is the distance between the quark and anti-quark in the $q\bar{q}$ pair, $\sigma=0.223$ GeV\textsuperscript{2}, and $\alpha=0.385$. We take $m_Q=4.7$ GeV for bottomonium. The  anisotropic Debye mass $\mu$ \cite{Krouppa:2015yoa} is defined as
\begin{eqnarray}
	\frac{\mu}{m_D}\equiv 1-\xi\frac{3+\cos{\theta}}{16},\nonumber
\end{eqnarray}
where $\xi$ is the anisotropic parameter. $\xi=0$ in an iso\-trop\-ic plasma. Hence we take $\mu=m_D$, where $m_D$ is the Debye mass,
\begin{eqnarray}\label{mD}
	m_D^2=(1.4)^2 N_c(1+N_f/6)4\pi\alpha_s T^2/3.
\end{eqnarray}
The factor of $(1.4)^2$ is an adjustment made to take into account higher-order corrections to the leading-order perturbative calculation of $m_D$ as determined using lattice calculations in \cite{Kaczmarek:2004gv}.

We will take $m_D\simeq 3\,p_{\mathrm{hard}}$, as was done in \cite{Krouppa:2015yoa}, where $p_{\mathrm{hard}}$ is a constant hard momentum scale for the particles in the QGP. From \cite{Margotta:2011ta}, we have that given a fixed number density, $p_{\mathrm{hard}}$ can be expressed as
\begin{eqnarray}
	p_{\mathrm{hard}}=(1+\xi)^{1/6}T.\nonumber
\end{eqnarray}
Hence we take $T=p_{\mathrm{hard}}$ for $\xi=0$. With parameters $N_c=3$, $N_f=2$ and $\alpha_s=3\alpha/4$ from the relation $\alpha\equiv C_F\alpha_s=[(N_c^2-1)/(2N_c)]\alpha_s$, then, Eq.~(\ref{mD}) yields $m_D^2=9.48\,T^2\simeq 9\,p_{\mathrm{hard}}^2$, as required.

\begin{figure}[!htbp]
	\subfloat[][]{
		\includegraphics[width=\columnwidth]{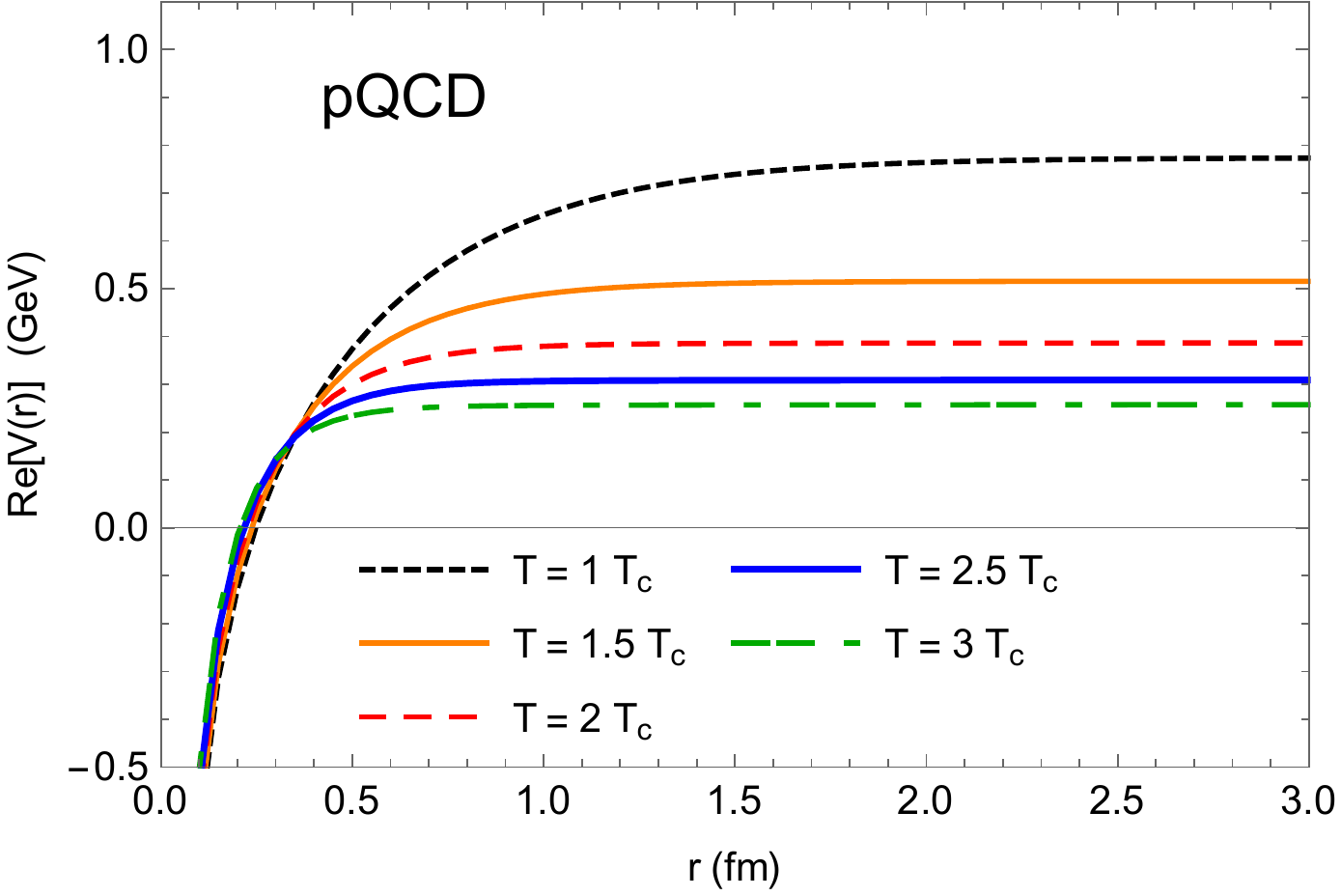}
		\label{fig:re_str_V}
	}\\
 	\subfloat[][]{
 		\includegraphics[width=\columnwidth]{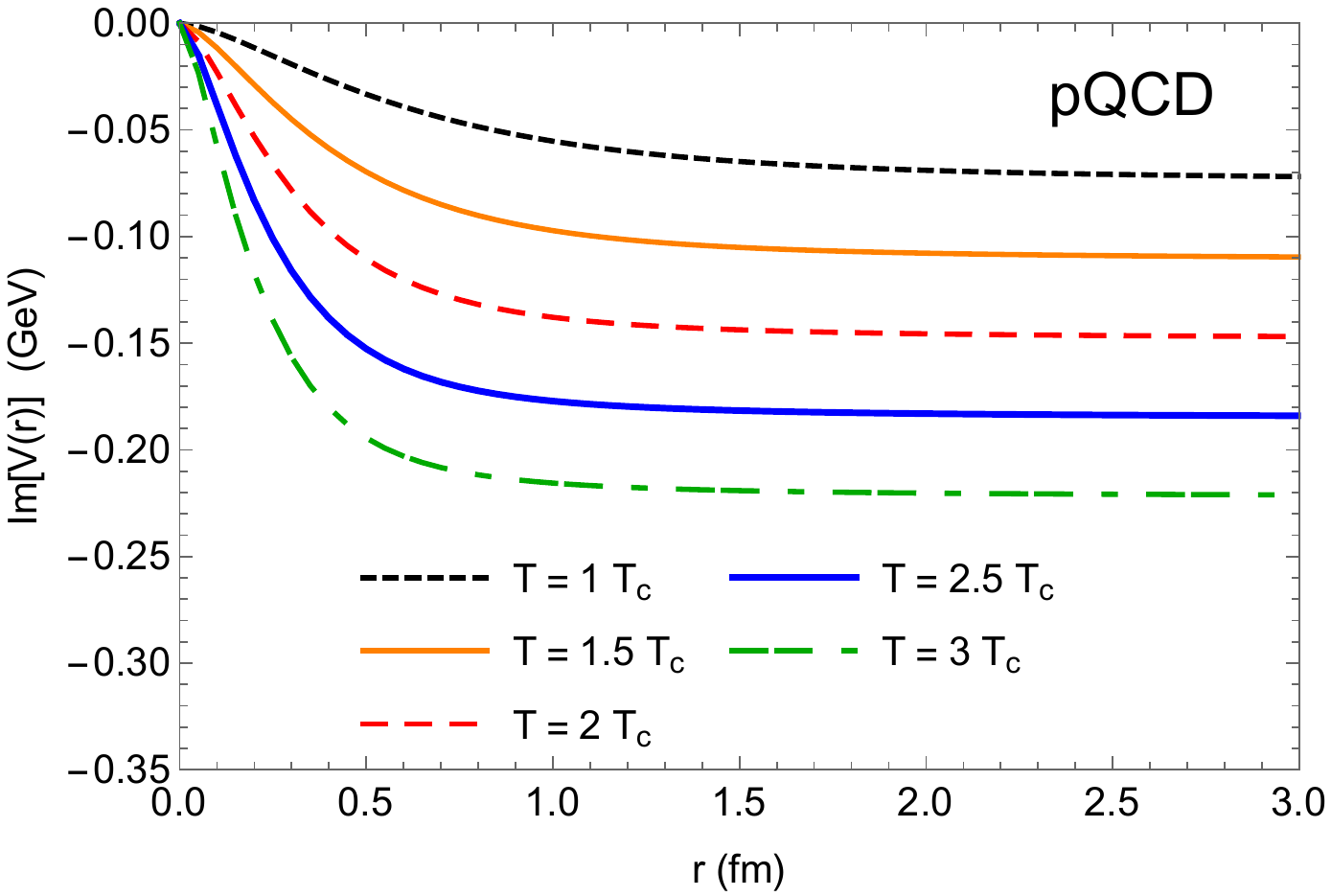}
 		\label{fig:im_str_V}
 	}
	\caption{\label{fig:str_V} Plot of the (\protect\subref*{fig:re_str_V}) real part of the potential $\Re[V(r)]$, given in Eq.\ (\protect\ref{ReVStr}), and the (\protect\subref*{fig:im_str_V}) imaginary part of the potential $\Im[V(r)]$, given in Eq.\ (\protect\ref{ImVStr}), as a function of the distance $r$ between the quark and anti-quark in the $b\bar{b}$, for various $T$ in an isotropic plasma. $T=T_c$ is denoted by the dotted black line, $T=1.5\ T_c$ by the solid orange line, $T=2\ T_c$ by the dashed red line, $T=2.5\ T_c$ by the thick blue line and $T=3\ T_c$ by the dashed-dotted green line. We take the critical temperature to be $T_c=192$~MeV \protect\cite{Margotta:2011ta}.
	}	
\end{figure}

We plot $\Re[V(r)]$ as a function of quark separation $r$ for various temperatures in Fig.~\ref{fig:re_str_V}.

The imaginary part of the weakly coupled potential is given by 
\begin{eqnarray}\label{ImVStr}
	\Im[V(r)]=\alpha T\{\phi(\hat{r})-\xi\left[\psi_1(\hat{r},\theta)+\psi_2(\hat{r},\theta)\right]\},
\end{eqnarray}
where $\hat{r}\equiv m_D r$ and
\begin{eqnarray}\label{phi}
	\phi(\hat{r})\equiv 2\int_0^\infty dz\frac{z}{(z^2+1)^2}\left[1-\frac{\sin{(z\hat{r})}}{z\hat{r}}\right]\!.
\end{eqnarray}
We do not include the functions $\psi_1(\hat{r},\theta)$ and $\psi_2(\hat{r},\theta)$ here since they are irrelevant in our isotropic plasma, but are available in \cite{Krouppa:2015yoa}.

We plot $\Im[V(r)]$ as a function of quark separation $r$ for various temperatures in Fig.~\ref{fig:im_str_V}. 

Note that, in order not to cut off the wave function prematurely, the potential must be considered at least as far as 20~fm. The function in Eq.\ (\ref{phi}) reduces in part to a Meijer G-function which becomes unstable for $r\gtrsim10$~fm. Hence, we imposed a maximum allowed value on $\hat{r}$ of $\hat{r}_{\mathrm{max}}=29.95$, which resulted in $\Im[V(r)]$ leveling off consistently as shown in Fig.~\ref{fig:im_str_V}.

\subsection{Strongly Coupled Quarkonia}\label{adscft}
We modeled the strongly coupled heavy quarkonia at rest in a QGP with the potential given in Albacete et al.\ \cite{Albacete:2008dz}. In that work, the authors derive the potential in $\mathcal{N}=4$ super Yang-Mills at finite temperature using AdS/CFT based on the methods of Rey et al.\ \cite{Rey:1998bq} and Brandhuber et al.\ \cite{Brandhuber:1998bs}.

The potential in \cite{Albacete:2008dz} is given by
\begin{eqnarray}\label{Vs}
	V_s(r)=\frac{\sqrt{\lambda}}{2c_0\pi}&&\left[-\frac{1}{z_{max}}\left(1-\frac{z_{max}^4}{z_h^4}\right)\right.\nonumber\\
	&&\times \left. F\left(\frac{1}{2},\frac{3}{4};\frac{1}{4};\frac{z_{max}^4}{z_h^4}\right)+\frac{1}{z_h}\vphantom{\frac{z_{max}^4}{z_h^4}}\right],
\end{eqnarray}
where $\lambda$ is the 't Hooft coupling, $c_0=\Gamma^2\left(\frac{1}{4}\right)/(2\pi)^{3/2}$, and $F$ is the usual Gaussian hypergeometric function. The temperature dependence comes from $z_h=1/\pi T$, and $z_{max}$ is found from the implicit equation
\begin{eqnarray}
	\label{eq:zmax}
	r c_0=\frac{z_{max}}{z_h^2}\sqrt{z_h^4-z_{max}^4}F\left(\frac{1}{2},\frac{3}{4};\frac{5}{4};\frac{z_{max}^4}{z_h^4}\right).
\end{eqnarray}

Note that $z_{\mathrm{max}}$, and hence the potential, becomes complex for $r>r_c\simeq0.870\,z_h$. The original papers \cite{Rey:1998bq,Brandhuber:1998bs} abandon the solution at this point, but \cite{Albacete:2008dz} does not. Considering that both weakly coupled pQCD and non-perturbative lattice QCD methods yield heavy quark potentials with both real and imaginary parts, it is sensible to expect the same using AdS/CFT methods. Further, allowing for complex $z_{\mathrm{max}}$ smooths out the kink in the potential from \cite{Rey:1998bq,Brandhuber:1998bs}, and does not violate any precepts of AdS/CFT.

\begin{figure}[!htbp]
	\subfloat[][]{
		\includegraphics[width=\columnwidth]{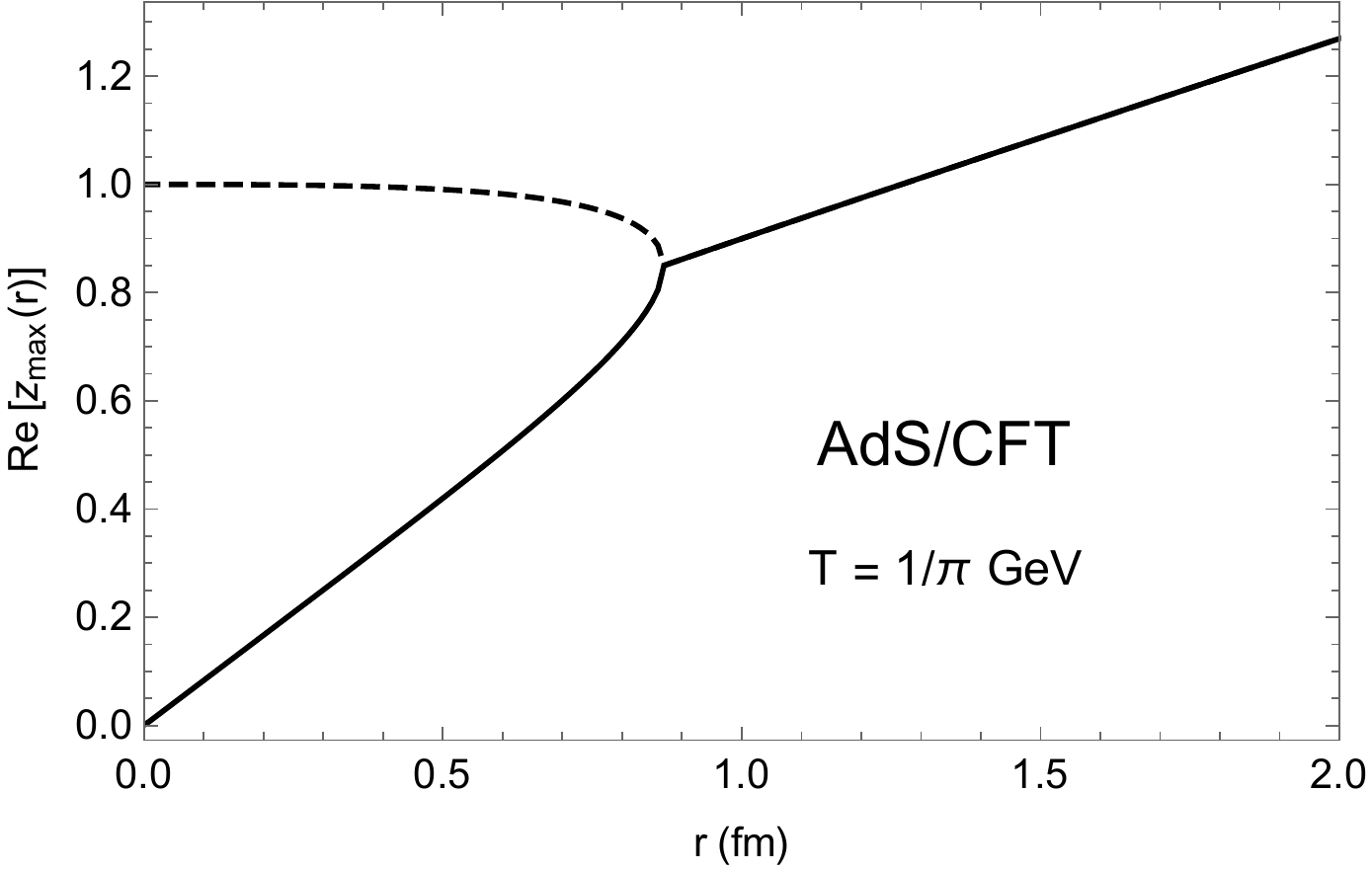}
		\label{fig:re_zmax_Re}
	}\\
 	\subfloat[][]{
 		\includegraphics[width=\columnwidth]{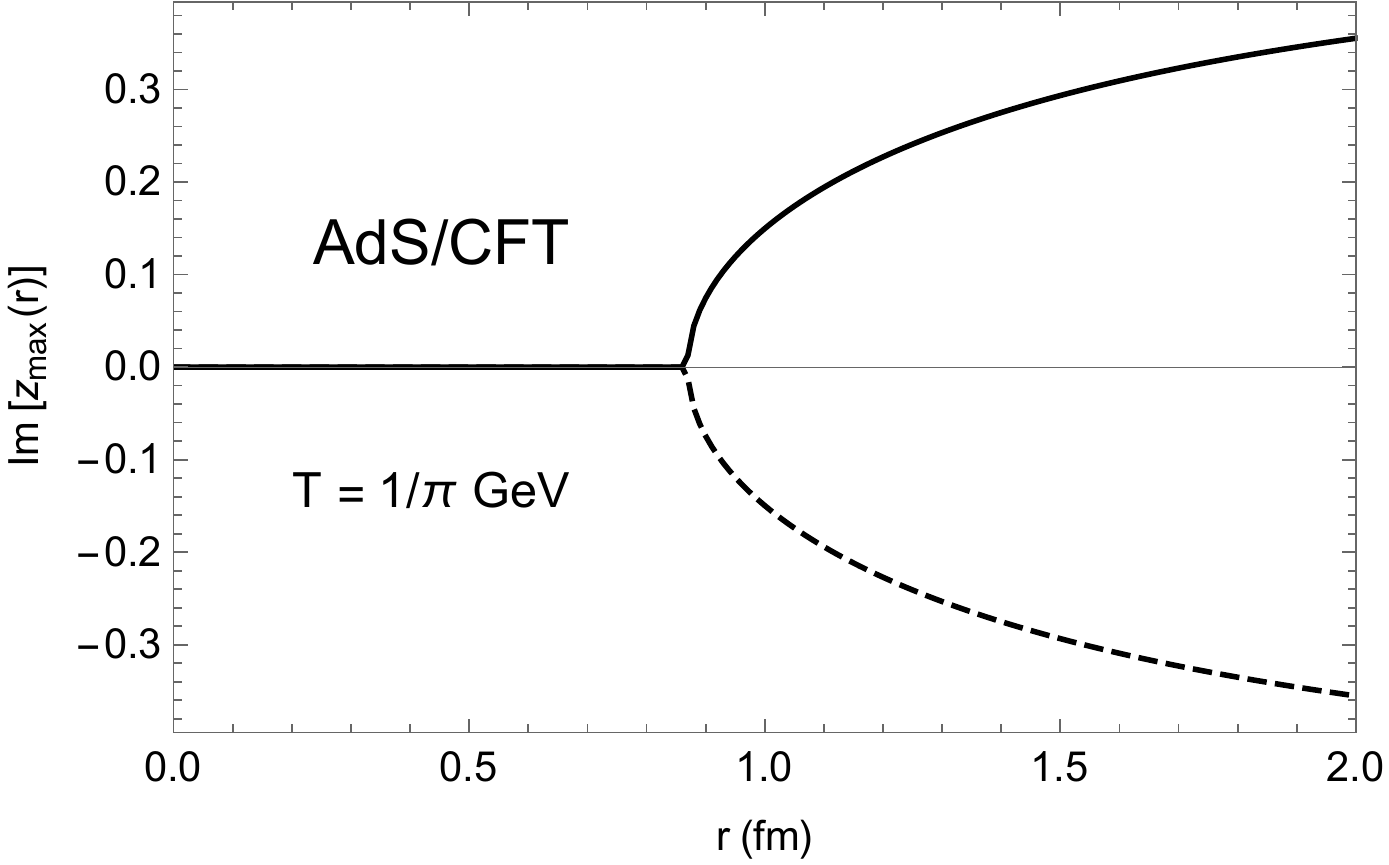}
 		\label{fig:re_zmax_Im}
 	}
	\caption{\label{fig:re_zmax} All possible solutions for (\protect\subref*{fig:re_zmax_Re}) $\Re[z_{\mathrm{max}}]$ and (\protect\subref*{fig:re_zmax_Im}) $\Im[z_{\mathrm{max}}]$, as a function of quark separation $r$ for $T=1/\pi$ GeV from Eq.\ (\protect\ref{eq:zmax}). The solid line is the root chosen over the dashed line solution for calculating the potential $V_s(r)$ as shown in Fig.~\ref{fig:alb_V}.}	
\end{figure}

Fig.~\ref{fig:re_zmax} shows the real and imaginary parts of $z_{\mathrm{max}}$, respectively, versus quark separation~$r$ for $T=1/\pi$~GeV.

The solution for $\Re[z_{\mathrm{max}}]$ denoted by the solid line is chosen as the physically relevant one, as was done in \cite{Albacete:2008dz}. This choice is made because that solution agrees with the solution first found by Maldacena \cite{Maldacena:1998im} for a heavy quark potential in vacuum for $\mathcal{N}=4$ SYM theory, given in Eq.\ (\ref{mald}), which reduces to $z_{\mathrm{max}}=rc_0$ at zero temperature.

Furthermore, the solid root is chosen for $\Im[z_{\mathrm{max}}]$ over its complex conjugate. This choice is justified as follows.  The time evolution of the wave function of any given state is $e^{-iEt}\sim e^{\Im[E]t}$. In order to ensure that the probability of a single state does not exceed one, we require that $\Im[E]<0$ \cite{Albacete:2008dz}, and therefore $\Im[V_s(r)]<0$. We thus choose the positive root for $\Im[z_{\mathrm{max}}]$, which yields the required negative $\Im[V_s(r)]$.

\begin{figure*}[!htbp]
	\subfloat[][]{
		\includegraphics[width=\columnwidth]{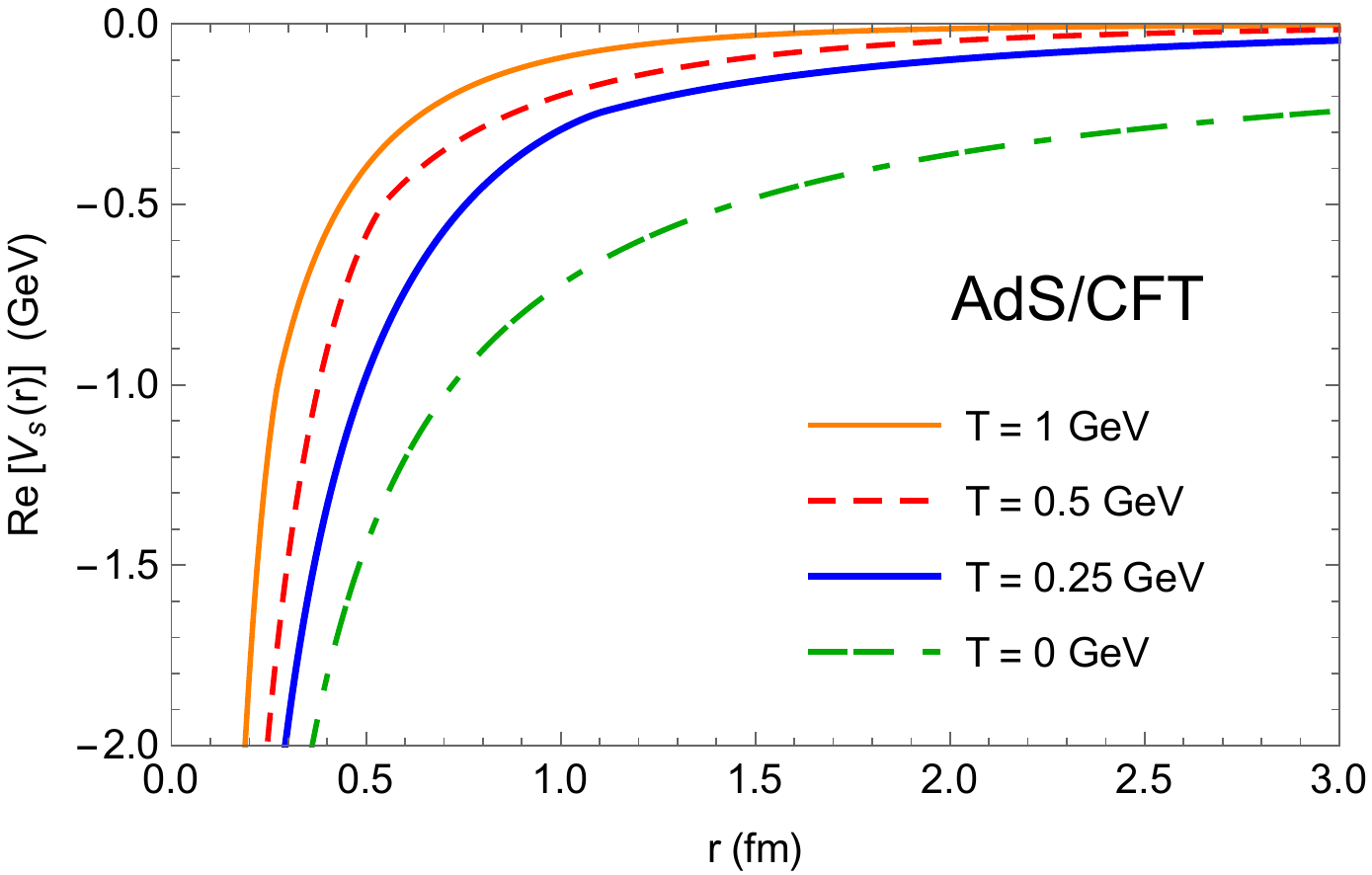}
		\label{fig:re_alb_V}
	}\ \ 
 	\subfloat[][]{
 		\includegraphics[width=\columnwidth]{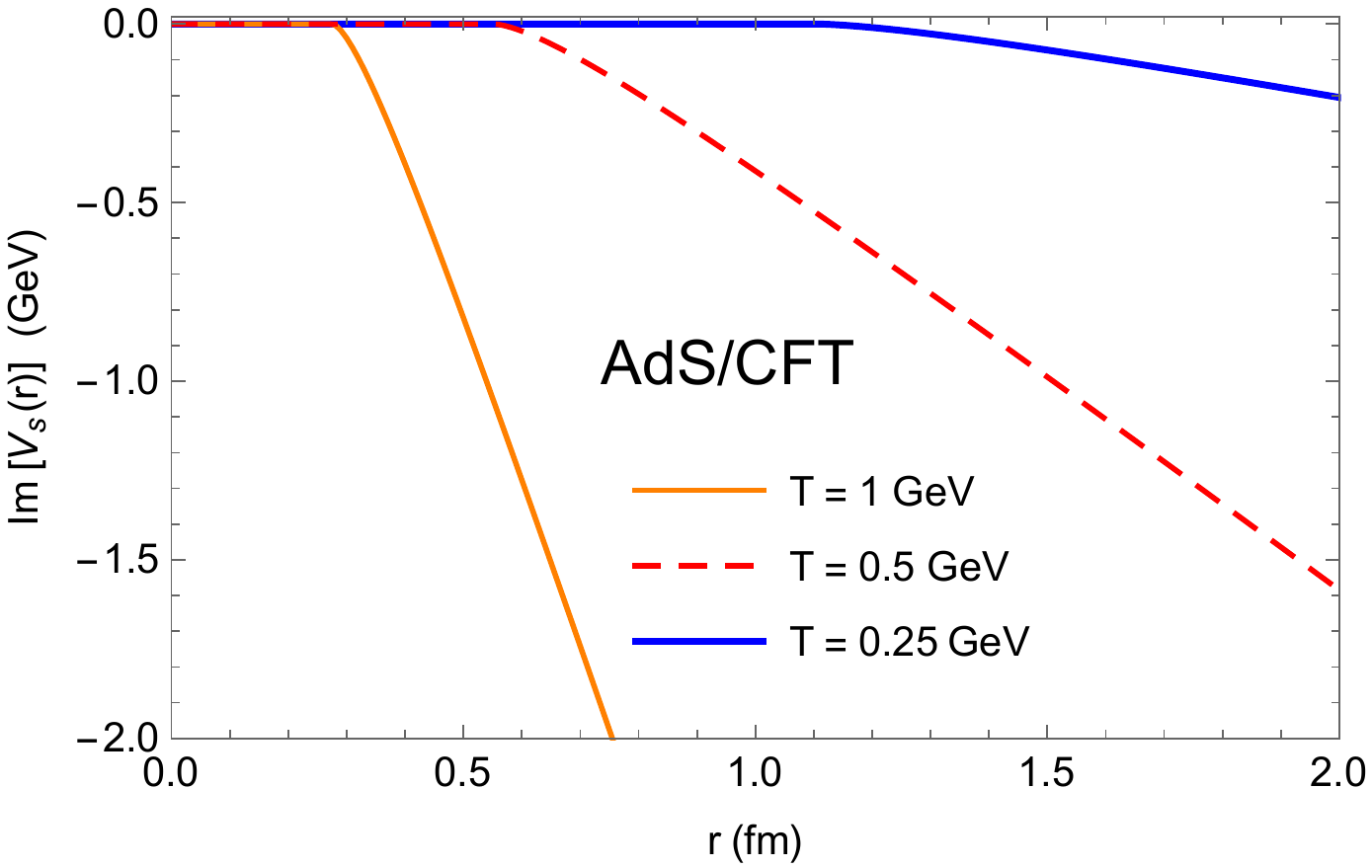}
 		\label{fig:im_alb_V}
 	}
	\caption{\label{fig:alb_V} The (\protect\subref*{fig:re_alb_V}) real part of the strongly coupled potential $\Re[V_s(r)]$ and the (\protect\subref*{fig:im_alb_V}) imaginary part of the potential $\Im[V_s(r)]$ from Eq.\ (\protect\ref{Vs}) as a function of the distance $r$ between the quark and anti-quark in the $b\bar{b}$, for various temperatures $T$ in an isotropic plasma. $T=0.25$~GeV is the thick blue line, $T=0.5$ GeV the dashed red line and $T=1$ GeV the thin orange line. $T=0$~GeV is included in (\protect\subref*{fig:re_alb_V}) as the dashed-dotted green line, taken from Eq.\ (\protect\ref{mald}).}	
\end{figure*}

Fig.~\ref{fig:re_alb_V} and \ref{fig:im_alb_V} show the real and imaginary parts of $V_s(r)$ as a function of quark separation $r$, calculated using the roots chosen for $z_{\mathrm{max}}$ as shown in Fig.~\ref{fig:re_zmax}, and taking $\lambda=10$.

As mentioned earlier, \cite{Maldacena:1998im} gives the heavy quark potential in a vacuum at zero temperature as
\begin{eqnarray}\label{mald}
	V_0(r)=-\frac{\sqrt{\lambda}}{2\pi c_0^2r}.
\end{eqnarray}
The plot of $V_0(r)$ is given in Fig.~\ref{fig:re_alb_V} for comparison.

Note that the real parts of the pQCD and AdS/CFT potentials, shown in Fig.~\ref{fig:re_str_V} and \ref{fig:re_alb_V}, respectively, are similar in form, but the imaginary parts from pQCD and AdS/CFT, shown in Fig.~\ref{fig:im_str_V} and \ref{fig:im_alb_V}, respectively, differ greatly: the imaginary part of the pQCD potential saturates as a function of $r$ whereas that of the AdS/CFT potential diverges.

\section{Numerical Integration of TDSE}\label{sec:TDSE}

The methodology used here follows that of \cite{Krouppa:2015yoa,Margotta:2011ta,1751-8121-40-8-013}, adapted to the special case of an isotropic plasma, with various modifications of the discretization as explained further below. In order to compute the ground state wave function, and hence the binding energy, we need to solve the non-relativistic, time dependent Schr\"{o}dinger Equation (TDSE) in spherical coordinates, subject to a spherically symmetric wave function $\Psi=\Psi(r,t)$. The TDSE is thus
\begin{equation}\label{tdse}
	i\partial_t \Psi(r,t)=H\Psi(r,t),
\end{equation}
where the Hamiltonian $H$ is
\begin{eqnarray}
	&&H=-\frac{1}{2m}\nabla^2+V(r),\nonumber\\
	&&\nabla^2=\frac{1}{r^2}\frac{\partial}{\partial r}\left(r^2\frac{\partial}{\partial r}\right)+f(\theta,\phi),
\end{eqnarray}
where $m\equiv m_1m_2/(m_1+m_2)$ is the reduced mass of the quarkonium. Note that $f(\theta,\phi)$ can be neglected since we are dealing with an isotropic quark gluon plasma. We take $\hbar=c=1$ throughout the paper.

Performing a Wick rotation to an imaginary time $\tau\equiv it$, Eq.\ (\ref{tdse}) has the general solution
\begin{eqnarray}
	\Psi(r,\tau)=\sum_{n=0}^\infty c_n\psi_n(r)e^{-E_n\tau}.
\end{eqnarray}

Since $E_n>E_0$ for all $n>0$, one can evolve forward in imaginary time such that all the higher order wave functions are suppressed and only the ground state wave function remains:
\begin{eqnarray}
	\lim_{\tau\rightarrow\infty} \Psi(r,t)\rightarrow c_0\psi_0(r)e^{-E_0\tau}\,,
\end{eqnarray}
where $\psi_0(r)$ is the ground state wave function and $E_0$ the ground state energy.

For simplicity, we will redefine the imaginary time to be dimensionless, $\tau\equiv itm$, along with further dimensionless quantities $\rho$, $R(\rho)$, and $W(\rho)$,
\begin{eqnarray}\label{dims}
	\rho\equiv mr\ ,\quad R(\rho,\tau)\equiv r\Psi(r,\tau)\ ,\quad W(\rho)\equiv\frac{V(\rho)}{m}.
\end{eqnarray}

The TDSE in terms of the dimensionless quantities is
\begin{equation}\label{dimtdse}
	\frac{\partial}{\partial\tau}R(\rho)=\frac{1}{2}\left(\frac{\partial^2}{\partial\rho^2}R(\rho)\right)-W(\rho)R(\rho).
\end{equation}

We improve upon the  finite difference time domain (FDTD) scheme implemented in \cite{Margotta:2011ta} (and elaborated upon in \cite{1751-8121-40-8-013}) by using a Crank-Nicolson Scheme, since this scheme is stable for much larger time steps $\Delta\tau$ \cite{Press:2007:NRE:1403886}. Given a PDE of the form
\begin{eqnarray}
	\frac{\partial u}{\partial t}=\frac{\partial}{\partial x}\left(D\frac{\partial u}{\partial x}\right)
\end{eqnarray}
one can discretize it as follows:
\begin{eqnarray}
	\frac{u_j^{n+1}-u_j^n}{\Delta t}=
	\frac{D}{2}&&\frac{(u_{j+1}^{n+1}-2u_j^{n+1}+u_{j-1}^{n+1})}{(\Delta x)^2}\nonumber\\
	&&+\frac{D}{2}\frac{(u_{j+1}^{n}-2u_j^{n}+u_{j-1}^{n})}{(\Delta x)^2},
\end{eqnarray}
where we use the notation
\begin{eqnarray}
	u_{j+1}^{n+1}\equiv u\left[(n+1)\Delta t,(j+1)\Delta x\right].\nonumber
\end{eqnarray}

Letting
\begin{eqnarray}
	W(\rho)R(\rho)=W(\rho_i)\left(\frac{R_{i}^{n}+R_{i}^{n+2}}{2}\right),
\end{eqnarray}
where we define $\tau_n\equiv(n-1)\Delta\tau$ and $\rho_i\equiv(i-1)\Delta\rho$,
the discretization of Eq.\ (\ref{dimtdse}) is then
\begin{widetext}
\begin{eqnarray}
	\left[-\frac{\Delta\tau}{4(\Delta\rho)^2}\right]R_{i+1}^{n+1}
	+\left[1+\frac{2\Delta\tau}{4(\Delta\rho)^2}+\frac{W(\rho_i)\Delta\tau}{2}\right]R_{i}^{n+1}
	+&&\left[-\frac{\Delta\tau}{4(\Delta\rho)^2}\right]R_{i-1}^{n+1}=\nonumber\\
	\left[\frac{\Delta\tau}{4(\Delta\rho)^2}\right]R_{i+1}^{n}&&
	+\left[1-\frac{2\Delta\tau}{4(\Delta\rho)^2}-\frac{W(\rho_i)\Delta\tau}{2}\right]R_{i}^{n}
	+\left[\frac{\Delta\tau}{4(\Delta\rho)^2}\right]R_{i-1}^{n}.
\end{eqnarray}
\end{widetext}

For the binding energy plots shown in Section~\ref{sec:RES}, we used $\Delta r=0.01$ fm and $\Delta\tau=10(\Delta\rho)^2$, where we have that $\Delta\rho=m\Delta r$. Note that for both the weakly and strongly coupled potentials, we implemented an $r_{\mathrm{cut}}=10^{-5}/m_Q>0$ on the potentials to ensure that solutions did not blow up at $r=0$.

\begin{figure*}[!tbp]
	\subfloat[][]{
		\includegraphics[width=.81\columnwidth]{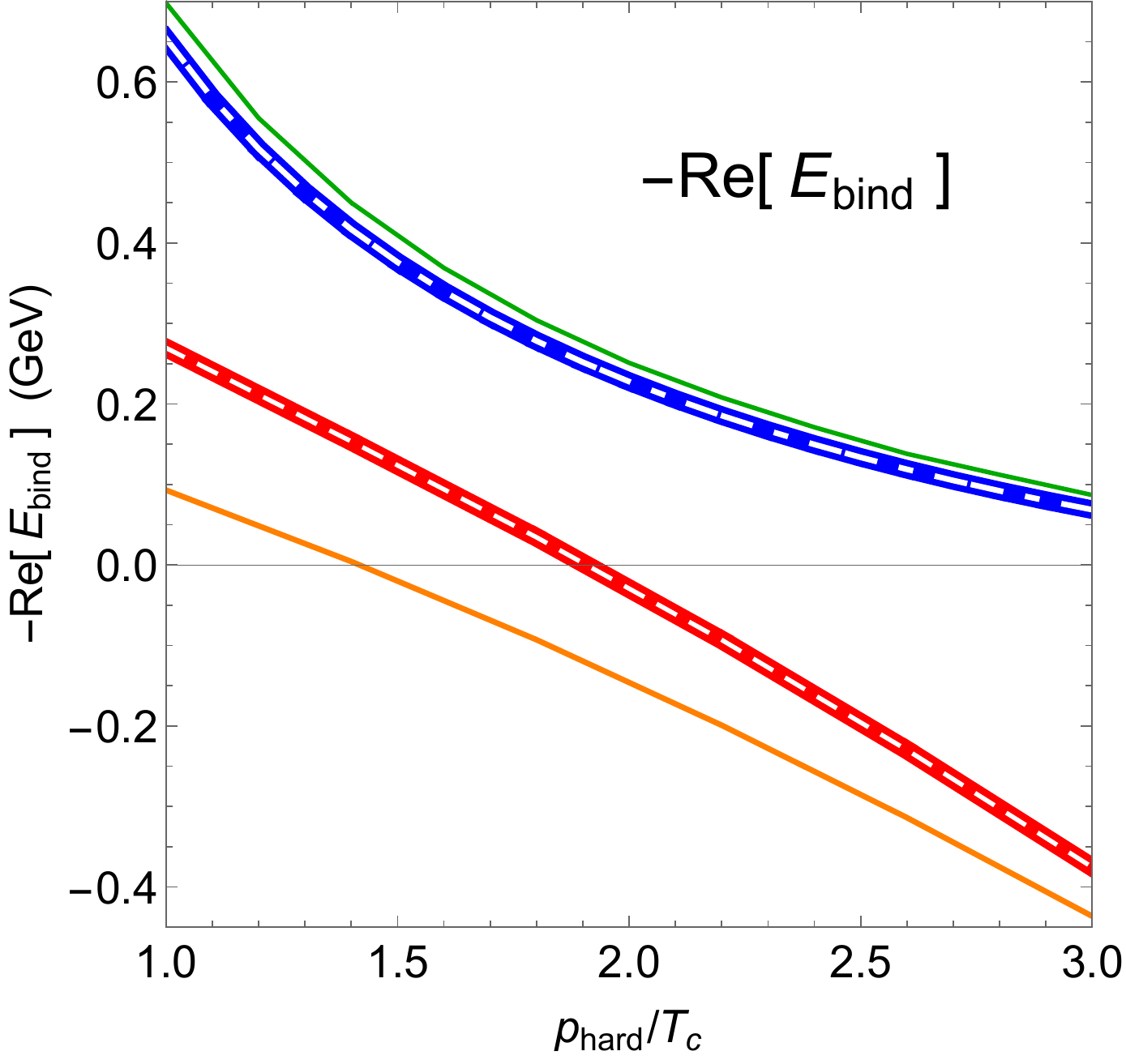}
		\label{fig:reebindall}
	}\ \ 
 	\subfloat[][]{
		\includegraphics[width=.79\columnwidth]{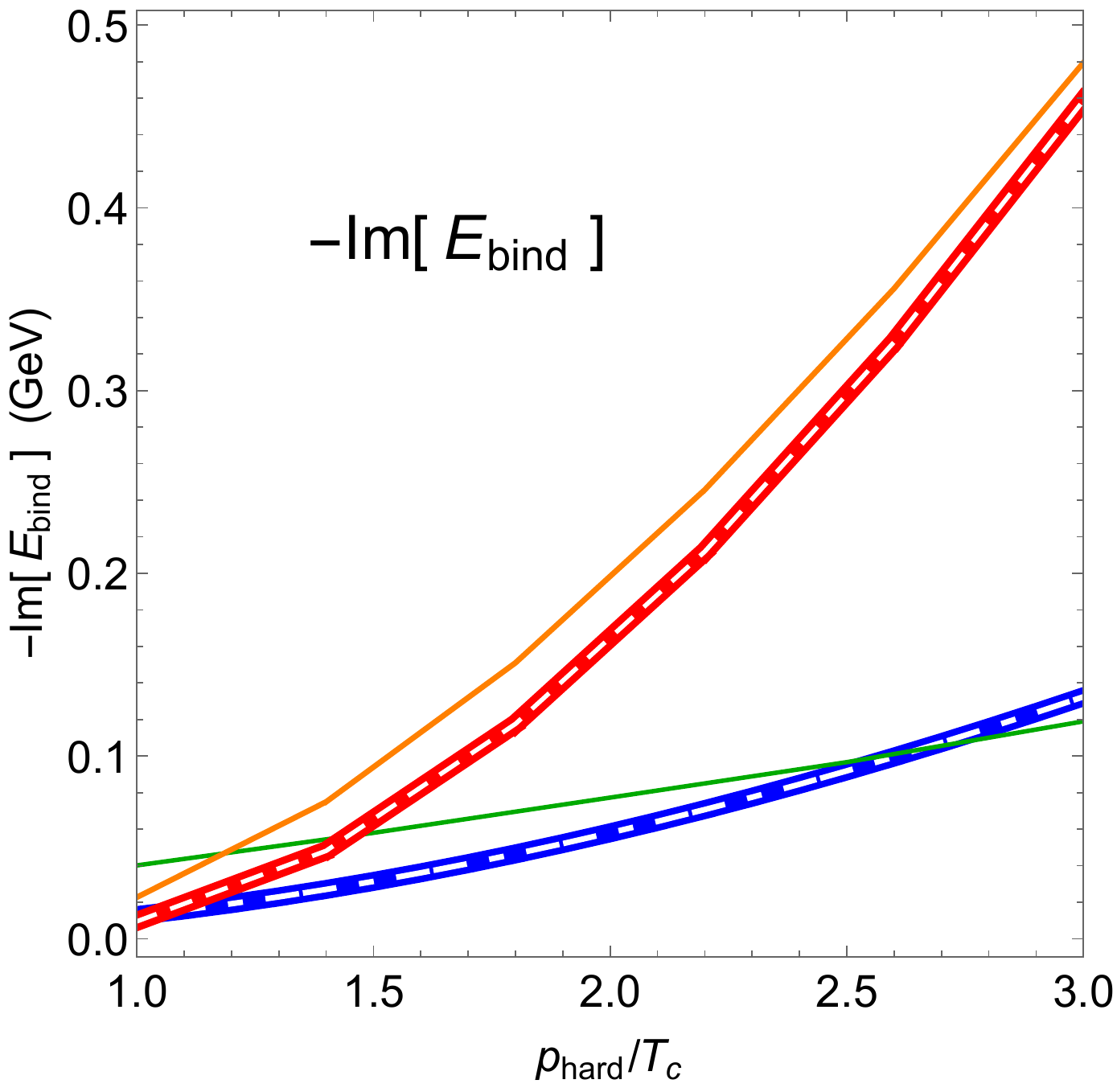}
 		\label{fig:imebindall}
 	}\ \ 
	\begin{minipage}[b]{.37\columnwidth}
 		\includegraphics[width=\columnwidth]{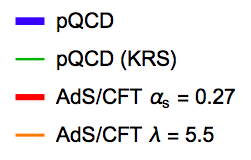}
		\vspace*{20pt}
 	\end{minipage}
	\caption{\label{fig:ebindall} The (\protect\subref*{fig:reebindall}) negative real part of $E_{\mathrm{bind}}$ and (\protect\subref*{fig:imebindall}) negative imaginary part of $E_{\mathrm{bind}}$ for $\Upsilon$(1S). The blue, red, and orange curves give the results for weakly coupled and strongly coupled ($\lambda=10$ and $\lambda=5.5$) $\Upsilon$(1S), respectively, computed from the imaginary time method of Section~\ref{sec:TDSE}. The dashed white curves inside the blue and red curves are from the independent evaluation using the complex variational method of Appendix~\ref{app:RITZ}. 
	The results from KRS \cite{Margotta:2011ta}, which should be identical to the blue curves, are given as solid green for comparison.}
\end{figure*}

The binding energy of the state can then be found from
\begin{eqnarray}\label{bind}
	E_{\mathrm{bind}}\equiv E_0-\Re[V(|r|\rightarrow\infty)],
\end{eqnarray}
where the ground state energy $E_0$ can be found from the ground state wave function,
\begin{eqnarray}\label{E0}
	E_0=\frac{\int r^2\,dr\,\psi_0(r)^*\,H\,\psi_0(r)}{\int r^2\,dr\,|\psi_0|^2}.
\end{eqnarray}

For weakly coupled quarkonia, the real part of the potential Eq.\ (\ref{ReVStr}) at infinity reduces to a constant,
\begin{eqnarray}
	\Re[V(|r|\rightarrow\infty)]=\frac{2\sigma}{\mu}.
\end{eqnarray}
In the case of the strongly coupled potential Eq.\ (\ref{Vs}), $\Re[V_s(|r|\rightarrow\infty)]=0$.

\section{Binding Energy Results}
\label{sec:RES}

Fig.~\ref{fig:reebindall} is a plot of the real part of the binding energy of $\Upsilon$(1S) from the pQCD potential, Eq.\ (\ref{ReVStr}) and (\ref{ImVStr}), and strongly coupled potential, Eq.\ (\ref{Vs}), as a function of temperature. Similarly, Fig.~\ref{fig:imebindall} gives the imaginary part of the binding energies for all cases mentioned above. 

For the AdS/CFT results, we show the binding energy both for the case where the coupling constant is $\lambda=10$ (labeled as $\alpha_s=0.27$) and where $\lambda=5.5$; the reasoning behind the choice of these values is explained in Section~\ref{sec:RAA}. The binding energy results for bottomonium from \cite{Margotta:2011ta} are labeled ``pQCD (KRS)'' and are included for comparison. 

Both the binding energy results presented for the pQCD potential and the AdS/CFT potential taking $\lambda=10$ were independently confirmed using a complex variational method, further explained in Appendix~\ref{app:RITZ}.

The binding energy found from our adapted methodology for the pQCD potential, Eq.\ (\ref{ReVStr}) and (\ref{ImVStr}), differs quantitatively from that presented in \cite{Margotta:2011ta}, which was used in Krouppa et al.\ \cite{Krouppa:2015yoa} to calculate suppression. In the case of $\Upsilon$(1S), this difference does not change the qualitative behavior of the quarkonia, since both results suggest that the quarkonia remain bound up to at least $T=3\,T_c$. However, we will see in Section~\ref{sec:RAA} that the small quantitative differences in the derived binding energies lead to a significant quantitative difference in the predicted suppression.

In the case of ground state charmonium, $J/\psi$, however, we see a qualitative difference in binding energies. The plot given in \cite{Margotta:2011ta} agrees with the commonly accepted wisdom from lattice QCD that $J/\psi$ mesons cease to exist as a bound state between $1.5\ T_c$ and $2.5\ T_c$.  However, using the same potential in \cite{Margotta:2011ta} but with our methodology, we find that the $J/\psi$ do not dissociate in this temperature range.

That our results for the pQCD potential from \cite{Margotta:2011ta} indicate bound charmonia above $T\gtrsim3T_c$ suggests a need either for an adjustment to the potential given in \cite{Krouppa:2015yoa}, since we believe our results to be more accurate than those presented in \cite{Margotta:2011ta}, or a slightly different approach in the interpretation of lattice QCD results with regards to the dissociation of $J/\psi$ at $T\sim1.5\ T_c$. We suspect that the plot presented in \cite{Margotta:2011ta} resulted from the authors integrating over too small a spatial region (they indicated that they considered an $r_{\mathrm{max}}$ of 5~fm, whereas our results were sensitive to values of $r_{\mathrm{max}}$ as large as $\sim$15~fm) or not imposing restrictions on the Meijer G-function present in the $\Im[V(r)]$ as we did in Section~\ref{Strick}.

The imaginary part of the binding energies from AdS/CFT are notably larger than those of weakly coupled quarkonia, and rise more steeply. This result is not surprising as the AdS/CFT potential has a divergent imaginary part, compared to the saturation of the imaginary part of the pQCD potential. 

Unlike in the case of the weakly coupled quarkonia where the $\Upsilon$(1S) remains bound for the temperature range considered, the strongly coupled $\Upsilon$(1S) dissociates at $\sim$1.9$\,T_c$. The comparatively larger imaginary part of the binding energy up to the temperature at which the bottomonium dissociates implies a much larger thermal width at higher $T$, and hence a larger suppression.

\begin{figure*}[!htbp]
	\subfloat[][]{
		\includegraphics[width=\columnwidth]{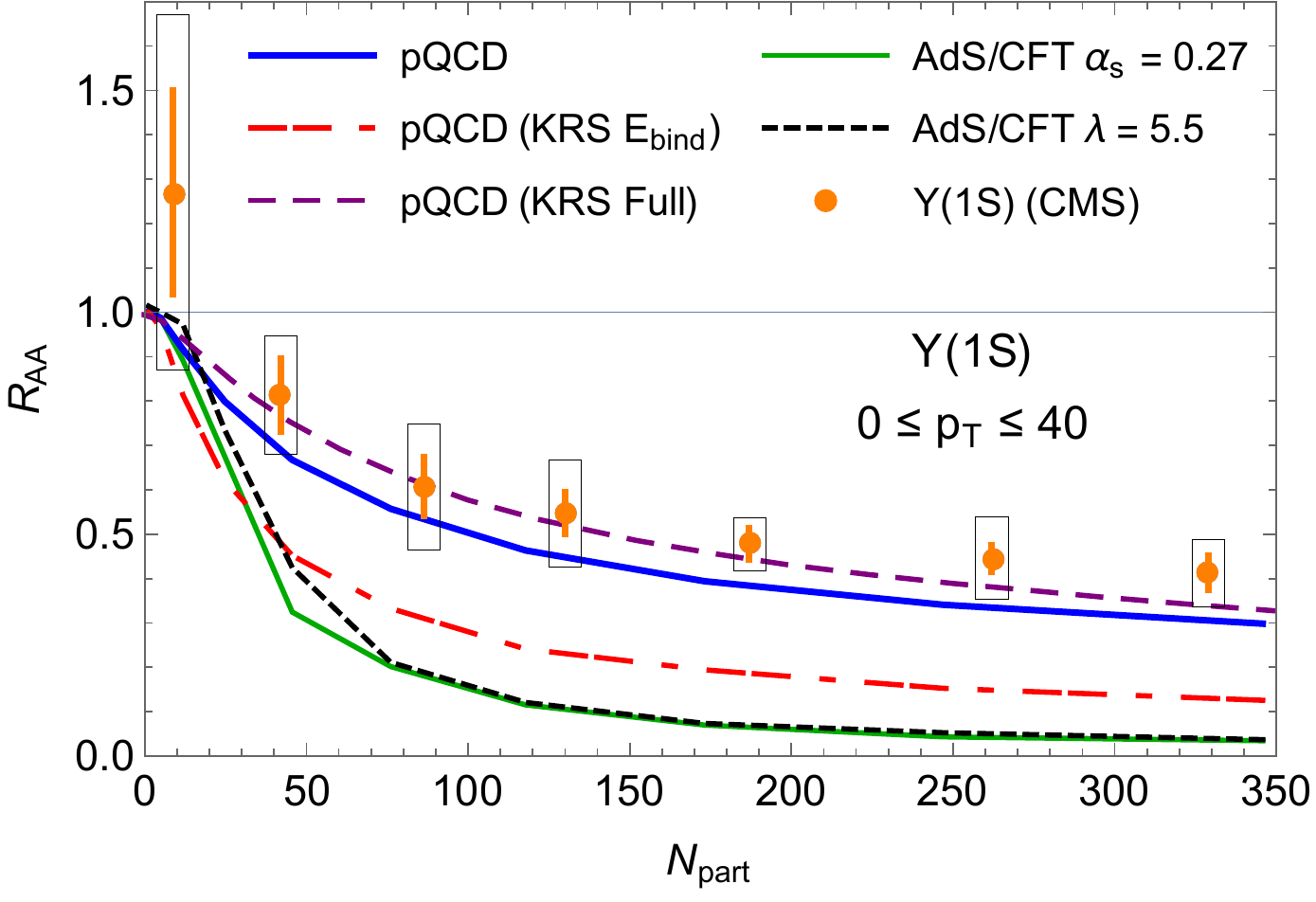}
		\label{fig:RAAvsNpart}
	}\ \ 
 	\subfloat[][]{
 		\includegraphics[width=\columnwidth]{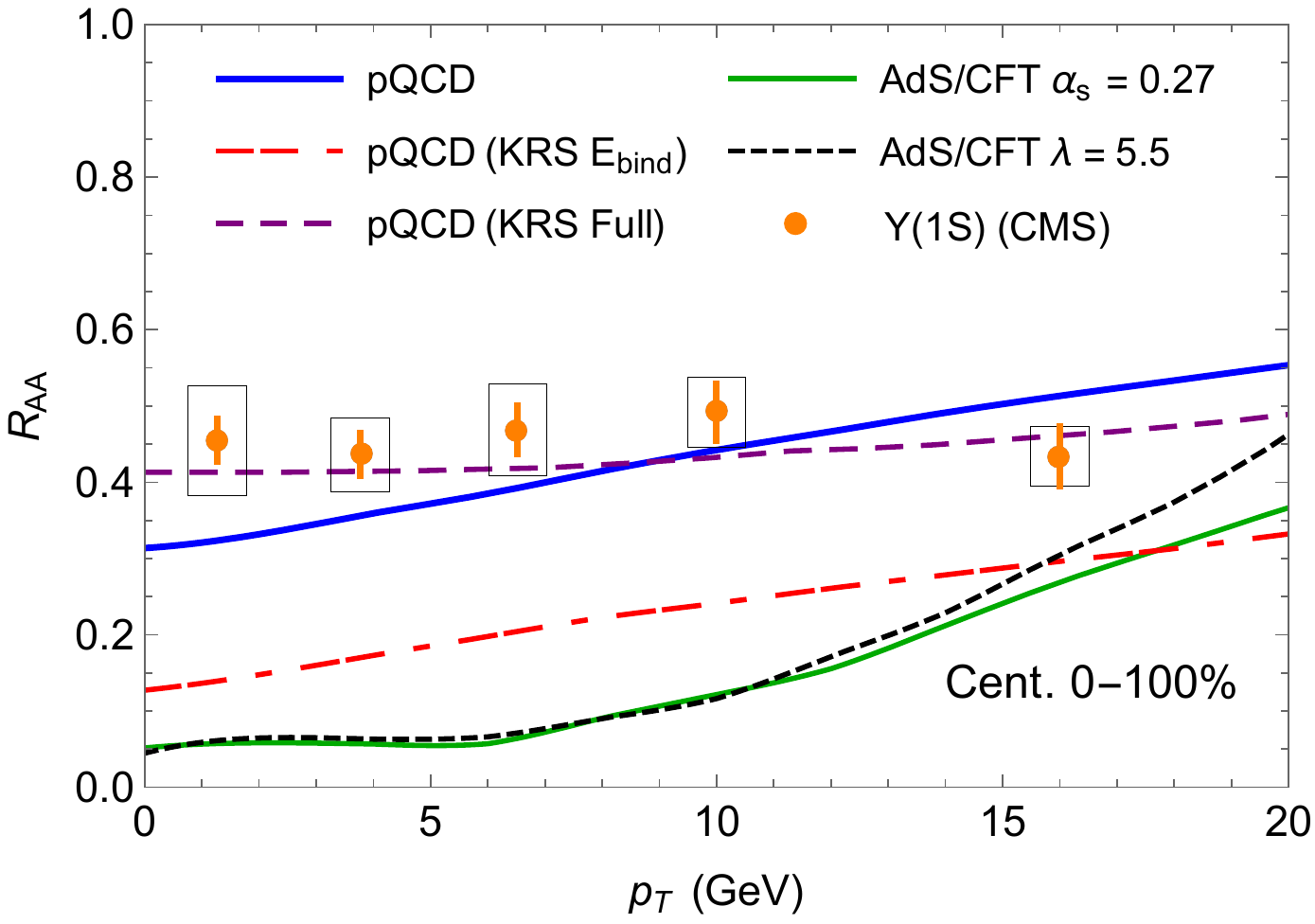}
 		\label{fig:RAAvspT}
 	}
	\caption{\label{fig:RAA} (\protect\subref*{fig:RAAvsNpart}) Nuclear modification factor $R_{AA}$ as a function of the number of participating nucleons $N_\mathrm{part}$ for $0\leq p_T\leq 40$. (\protect\subref*{fig:RAAvspT})~Nuclear modification factor $R_{AA}$ as a function of transverse momentum $p_T$ for combined centrality classes. In both subfigures, the thick solid blue line gives our results for weakly coupled $\Upsilon$(1S), and the dashed-dotted red line that calculated for the binding energy from KRS \cite{Margotta:2011ta} using our suppression model. The $R_{AA}$ presented in KRS \cite{Krouppa:2015yoa} as calculated using their suppression model is given in dashed purple. The solid green and dotted black lines give the results for strongly coupled $\Upsilon$(1S) with coupling constants $\alpha_s=0.27$ (and $T_{SYM}=T_{QCD}$) and $\lambda=5.5$ (and $T_{SYM}=T_{QCD}/3^{1/4}$), respectively. Data from CMS \cite{Khachatryan:2016xxp} is included in orange.}	
\end{figure*}

\section{Suppression}\label{sec:RAA}

We would like to make quantitative predictions for the suppression of bottomonia in heavy ion collisions and compare to measured data.  
The nuclear modification factor $R_{AA}$ is calculated following \cite{Krouppa:2015yoa}:
\begin{eqnarray}
	&&R_{AA}(p_T,y,\mathbf{x}_\perp ,b)=e^{-\zeta(p_T,y,\mathbf{x}_\perp ,b)},\\
	&&\zeta\equiv\Theta(\tau_f-\tau_{\mathrm{form}})\int_{\max{(\tau_{\mathrm{form}},\tau_0)}}^{\tau_f} d\tau\, 		\Gamma(\tau,\mathbf{x}_\perp,\varsigma=y), \nonumber
\end{eqnarray}
where the thermal width $\Gamma(\tau,\mathbf{x}_\perp,\varsigma)$ is given as
\begin{eqnarray}
	\Gamma(\tau,\mathbf{x}_\perp,\varsigma)=
	\begin{cases}
		-2\Im[E_{bind}] & \Re[E_{bind}]<0 \\
		\gamma_{\mathrm{dis}}	& \Re[E_{bind}]\geq 0.
	\end{cases}
\end{eqnarray}
We take $\gamma_{\mathrm{dis}}=10$ GeV, as was done in \cite{Krouppa:2015yoa}. Furthermore, $b$ is the impact parameter, and $y$ the rapidity, taken to be zero. The formation time $\tau_\mathrm{form}$ is calculated using
\begin{eqnarray}
	\tau_\mathrm{form}=E_T\,\tau_\mathrm{form}^0/m_Q
\end{eqnarray}
where $\tau_\mathrm{form}^0=0.2$ fm is taken for the initial formation time of the state \cite{Krouppa:2015yoa}. Lastly, the final time $\tau_f$ is taken as the time at which the temperature $T$ of the QGP drops below the critical temperature $T_c$.

We use the optical limit of the Glauber model \cite{Miller:2007ri} to describe the background in the case of $\sqrt{s_\textrm{NN}}=2.76$~TeV Pb+Pb collisions. Taking a weighted average over the region with limits $\mathbf{x}_\perp=[-10, 10]$ fm, we have
\begin{equation}\label{raaptb}
	R_{AA}(p_T,b)=\frac{\int d^2\mathbf{x}_\perp d\phi\;T_\mathrm{AA}(\mathbf{x}_\perp,b)\;
		R_{AA}(p_T,y,\mathbf{x}_\perp ,b)}{2\pi N_{\mathrm{coll}}},
\end{equation}
where $T_\mathrm{AA}(\mathbf{x}_\perp,b)$ is the nuclear overlap function and $N_{\mathrm{coll}}\equiv\int d^2\mathbf{x}_\perp\; T_\mathrm{AA}(\mathbf{x}_\perp,b)$ is the number of binary nu\-cle\-on-nucleon collisions in the region.

We set a central temperature $T_0=522$ MeV and initial formation time $\tau_0=0.3$ fm, as is done in \cite{Krouppa:2015yoa}. Hence the temperature $T$ of the QGP is given by
\begin{eqnarray}
	T(\tau_0,\mathbf{x}_\perp,b)=\beta\left(\frac{\rho_\mathrm{part}(\tau_0,\mathbf{x}_\perp,b)}{\tau_0}\right)^{1/3}
\end{eqnarray}
where $\beta=0.231$ GeV fm is a proportionality constant and $\rho_\mathrm{part}$ the participant density.

Fig.~\ref{fig:RAAvsNpart} gives the nuclear modification factor $R_{AA}$ for each of the sets of binding energies shown in Fig.~\ref{fig:reebindall} and \ref{fig:imebindall} as a function of the number of participating nucleons $N_\mathrm{part}$. To this end, the $R_{AA}(p_T,b)$ from Eq.\ (\ref{raaptb}) is averaged over the transverse momentum range $0\leq p_T\leq 40$ with a weighting of $E^{-4}$ \cite{Krouppa:2015yoa}. 

Fig.~\ref{fig:RAAvspT} shows $R_{AA}(p_T)$, where all centrality classes are included, weighed by the number of binary nucleon-nucleon collisions $N_\mathrm{coll}$.

Suppression results for mid-rapidity ($|y|<2.4$) Pb+Pb collisions at $\sqrt{s_\textrm{NN}}=2.76$~TeV from the CMS Collaboration \cite{Khachatryan:2016xxp} are included in Fig.~\ref{fig:RAAvsNpart} and \ref{fig:RAAvspT} for comparison.  

We show in Fig.~\ref{fig:RAAvsNpart} and \ref{fig:RAAvspT} two predictions for the suppression of strongly coupled bottomonia in an attempt to at least partially map out some of the systematic theoretical uncertainties associated with the use of the AdS/CFT correspondence.  Since we used a potential derived in AdS-space dual to maximally supersymmetric Yang-Mills theory, there is no single obvious map between the parameters of QCD and of $\mathcal{N}=4$ SYM.  For the $\alpha_s=0.27$ curve, we took $\lambda_{SYM}=10$ and $T_{SYM}=T_{QCD}$, where $\lambda_{SYM}=10=\sqrt{4\pi\alpha_sN_c}$  (and thus $\alpha_s=0.27$ for $N_c=3$) is approximately the value of the QCD running coupling constant evaluated at the first Matsubara frequency of the plasma.  For the $\lambda=5.5$ curve, the coupling constant was set by a comparison to the $q\bar{q}$ potential from lattice and $T_{SYM}=T_{QCD}/3^{-1/4}$ is a result of assuming the entropies of the QCD and SYM plasmas are the same \cite{Gubser:2006nz}.

We show in Fig.~\ref{fig:RAAvsNpart} and \ref{fig:RAAvspT} three predictions for weakly coupled bottomonia: 1) the suppression using the binding energies we compute from the potential in \cite{Margotta:2011ta} run through our medium background, 2) the suppression using the binding energies computed in \cite{Margotta:2011ta} run through our medium background, and 3) the suppression quoted in \cite{Krouppa:2015yoa} in which they run the binding energies computed in \cite{Margotta:2011ta} through their background.

\section{Discussion and Outlook}\label{sec:DISC}
In this paper we computed for the first time the suppression of bottomonia in an isotropic strongly coupled QGP and compared the results to those from a weakly coupled QGP and to data from the CMS Collaboration \cite{Khachatryan:2016xxp}.

The non-relativistic, time dependent, radially symmetric Schr\"{o}dinger Equation was solved numerically in order to find the ground state wave functions for two potential models: one from pQCD and one from AdS/CFT. The discretized, numerical evaluation of the imaginary time Schr\"odinger Equation was performed using a Crank-Nicolson Scheme, evolving forward in imaginary time until all higher order wave functions were sufficiently suppressed. The potential used for weakly coupled quarkonia was taken from \cite{Margotta:2011ta}, in which the potential came from leading-order pQCD with various corrections. The strongly coupled quarkonia potential was taken from \cite{Albacete:2008dz}, who obtained their potential from AdS/CFT.

The ground state wave functions obtained were then used to find the (complex) ground state energies for $\Upsilon$(1S).  These ground state energies were then independently confirmed using a complex variational technique \cite{Kraft2013}. Our binding energies for the weakly coupled potential in \cite{Margotta:2011ta} differed somewhat from those found in \cite{Margotta:2011ta}, likely due to extending the physical region under consideration and from a possibly more careful treatment of the potential. For $\Upsilon$(1S), the difference was only quantitative, but for $J/\psi$ the difference was qualitative: for the potential in \cite{Margotta:2011ta}, we found charmonia remain bound up to at least $T\sim3 T_c$. 

Our first results for $\Upsilon$(1S) strongly coupled to a strongly coupled plasma show binding energies with much larger imaginary parts than those found from the pQCD potential, as well as real parts that become positive within the $T_c$ to $3\,T_c$ range considered.  Thus, for the potential models considered here, a strongly coupled $\Upsilon$(1S) interacting with a strongly coupled plasma melts at a \emph{lower temperature} than a weakly coupled $\Upsilon$(1S) interacting with a weakly coupled plasma. The $\Upsilon$(1S) hence appears more strongly bound at weak coupling than at strong coupling, which is surprising.  

Since the weak coupling bottomonia become more strongly bound as the coupling is increased and the strong coupling bottomonia become less strongly bound as the coupling decreases, that the weak coupling bottomonia is more strongly bound at weak coupling than at strong coupling suggests some non-monotonic behavior of the binding energies at the threshold between the weak and strong coupling regimes.  This non-monotonic behavior possibly stems from deriving the potential at weak coupling in QCD whereas the strong coupling potential was derived in the slightly different theory, $\mathcal{N}=4$ SYM; it would be interesting to compare binding energies from the quarkonium potential at weak and strong coupling consistently within $\mathcal{N}=4$ SYM.

We then input the complex ground state binding energies we found into an implementation of the suppression model described in \cite{Krouppa:2015yoa} to determine the $\Upsilon$(1S) nuclear modification factor $R_{AA}$ as a function of the number of participating nucleons, $N_\mathrm{part}$, and of transverse momentum, $p_T$, respectively.  The difference in binding energies for the two coupling scenarios is echoed in the $R_{AA}$ results: from the larger imaginary parts of the strongly coupled binding energies, we see a significantly larger suppression for strongly coupled $\Upsilon$(1S) than for weakly coupled $\Upsilon$(1S).  Quantitatively, our full model---comprised of the potential, the resulting quarkonia binding energies, and the translation to $R_{AA}$---significantly overpredicts the suppression of strongly coupled bottomonia compared to data. At the same time, our predictions for weakly coupled bottomonia are consistent with data.

We note that our model for the medium is significantly less sophisticated compared to that used in \cite{Krouppa:2015yoa}: our background is an optical Glauber model as opposed to the 3+1D viscous anisotropic hydrodynamics in that work. Our medium incorporates only Bjorken expansion, whereas the background in \cite{Krouppa:2015yoa} includes transverse expansion and entropy production. Therefore the plasma in \cite{Krouppa:2015yoa} cools faster than ours, leading to our model showing more dissociation for the same binding energies. The extent of the sensitivity of $R_{AA}$ to the background used is surprisingly large. With the only difference being the background geometry used, we ran the binding energies from \cite{Margotta:2011ta} through our suppression model and found an $R_{AA}$ a factor of two smaller than that shown in \cite{Krouppa:2015yoa}.

In contrast to the favorable comparison between the pQCD-based results of \cite{Krouppa:2015yoa} and the CMS data \cite{Chatrchyan:2012lxa}, if we assume our weak coupling binding energies are more accurate than those of \cite{Margotta:2011ta}, then computing $R_{AA}$ with the more sophisticated background from \cite{Krouppa:2015yoa} would likely yield a significant underprediction of the suppression of bottomonia. 

At strong coupling, with a potential derived from AdS/CFT as described in \cite{Albacete:2008dz}, it seems unlikely that the use of a more sophisticated background would reduce the suppression of bottomonia enough that the predicted $R_{AA}$ would be consistent with data; however, the differences from using a more sophisticated background, suppression model, and velocity dependent potential may ultimately be sufficient for future strongly coupled quarkonia predictions to be consistent with current data.  

We leave the implementation of more advanced calculations of quarkonia suppression---including better modeling of the medium background, more accurate initial quarkonia production, a more realistic dissociation model, and the use of velocity dependent potentials---and a more thorough investigation of systematic theoretical uncertainties in quarkonia $R_{AA}$ for future work.

\acknowledgments
The authors wish to thank the South African National Research Foundation and the SA-CERN Collaboration for their financial support.  The authors also wish to thank Michael Strickland for useful discussions.

\appendix

\section{Complex Variational Method}\label{app:RITZ}
The complex binding energies presented in Section~\ref{sec:RES} for the pQCD potential, as well as the AdS/CFT potential with a coupling constant of $\lambda=10$, were confirmed using the complex variational method presented here. We used the complex variational principle from \cite{Kraft2013}, which can be seen for full theorems and proofs, and is an extension of the usual Ritz variational method \cite{Messiah2014} to complex-valued potentials.

First, we define the c-product as
\begin{eqnarray}
	(\psi |\phi)=\int_{\mathbb{R}^n}\psi(\vec{x})\phi(\vec{x})\,d^nx
\end{eqnarray}
for two general functions $\psi(\vec{x})$ and $\phi(\vec{x})$. Then, given some eigenvalue problem $H\psi(\vec{x})=E\psi(\vec{x})$ where $(\psi|\psi)\neq 0$, we can define the Rayleigh quotient as
\begin{eqnarray}
	R(\vec{\alpha})\equiv\frac{(\psi|\hat{H}|\psi)}{(\psi|\psi)},
\end{eqnarray}
where $\psi(\vec{x};\vec{\alpha})$ is a parameter-dependent trial wave function, with $\vec{\alpha}\in\mathbb{C}^m$.

In the event that $H\psi(\vec{x};\vec{\alpha}_0)=E\psi(\vec{x};\vec{\alpha}_0)$ is satisfied for some $\psi(\vec{x};\vec{\alpha}_0)$ and $(\psi(\vec{\alpha}_0)|\psi(\vec{\alpha}_0))\neq 0$, then we have that
\begin{eqnarray}
	\frac{\partial R(\vec{\alpha}_0)}{\partial \alpha^i_0}=0\ .
\end{eqnarray}

Note that the complex variational method only guarantees that $\psi(\vec{x};\vec{\alpha}_0)$ is a stationary wave function of the eigenvalue problem, not necessarily the ground state, as is the case with the real-valued potential Ritz variational method. Since the imaginary time techniques described in Section~\ref{sec:TDSE} guarantee a ground state wave function if the imaginary time evolution is large enough, then a conclusion from these complex variational methods of the existence of a stationary state with binding energies equal to those from the imaginary time evolution is sufficient to confirm the imaginary time evolution ground state binding energy results.


\begin{thebibliography}{83}%
\makeatletter
\providecommand \@ifxundefined [1]{%
 \@ifx{#1\undefined}
}%
\providecommand \@ifnum [1]{%
 \ifnum #1\expandafter \@firstoftwo
 \else \expandafter \@secondoftwo
 \fi
}%
\providecommand \@ifx [1]{%
 \ifx #1\expandafter \@firstoftwo
 \else \expandafter \@secondoftwo
 \fi
}%
\providecommand \natexlab [1]{#1}%
\providecommand \enquote  [1]{``#1''}%
\providecommand \bibnamefont  [1]{#1}%
\providecommand \bibfnamefont [1]{#1}%
\providecommand \citenamefont [1]{#1}%
\providecommand \href@noop [0]{\@secondoftwo}%
\providecommand \href [0]{\begingroup \@sanitize@url \@href}%
\providecommand \@href[1]{\@@startlink{#1}\@@href}%
\providecommand \@@href[1]{\endgroup#1\@@endlink}%
\providecommand \@sanitize@url [0]{\catcode `\\12\catcode `\$12\catcode
  `\&12\catcode `\#12\catcode `\^12\catcode `\_12\catcode `\%12\relax}%
\providecommand \@@startlink[1]{}%
\providecommand \@@endlink[0]{}%
\providecommand \url  [0]{\begingroup\@sanitize@url \@url }%
\providecommand \@url [1]{\endgroup\@href {#1}{\urlprefix }}%
\providecommand \urlprefix  [0]{URL }%
\providecommand \Eprint [0]{\href }%
\providecommand \doibase [0]{http://dx.doi.org/}%
\providecommand \selectlanguage [0]{\@gobble}%
\providecommand \bibinfo  [0]{\@secondoftwo}%
\providecommand \bibfield  [0]{\@secondoftwo}%
\providecommand \translation [1]{[#1]}%
\providecommand \BibitemOpen [0]{}%
\providecommand \bibitemStop [0]{}%
\providecommand \bibitemNoStop [0]{.\EOS\space}%
\providecommand \EOS [0]{\spacefactor3000\relax}%
\providecommand \BibitemShut  [1]{\csname bibitem#1\endcsname}%
\let\auto@bib@innerbib\@empty
\bibitem [{\citenamefont {Adams}\ \emph {et~al.}(2005)\citenamefont {Adams}
  \emph {et~al.}}]{Adams:2005dq}%
  \BibitemOpen
  \bibfield  {author} {\bibinfo {author} {\bibfnamefont {John}\ \bibnamefont
  {Adams}} \emph {et~al.} (\bibinfo {collaboration} {STAR}),\ }\bibfield
  {title} {\enquote {\bibinfo {title} {{Experimental and theoretical challenges
  in the search for the quark gluon plasma: The STAR Collaboration's critical
  assessment of the evidence from RHIC collisions}},}\ }\href {\doibase
  10.1016/j.nuclphysa.2005.03.085} {\bibfield  {journal} {\bibinfo  {journal}
  {Nucl. Phys.}\ }\textbf {\bibinfo {volume} {A757}},\ \bibinfo {pages}
  {102--183} (\bibinfo {year} {2005})},\ \Eprint
  {http://arxiv.org/abs/nucl-ex/0501009} {arXiv:nucl-ex/0501009 [nucl-ex]}
  \BibitemShut {NoStop}%
\bibitem [{\citenamefont {Adcox}\ \emph {et~al.}(2005)\citenamefont {Adcox}
  \emph {et~al.}}]{Adcox:2004mh}%
  \BibitemOpen
  \bibfield  {author} {\bibinfo {author} {\bibfnamefont {K.}~\bibnamefont
  {Adcox}} \emph {et~al.} (\bibinfo {collaboration} {PHENIX}),\ }\bibfield
  {title} {\enquote {\bibinfo {title} {{Formation of dense partonic matter in
  relativistic nucleus-nucleus collisions at RHIC: Experimental evaluation by
  the PHENIX collaboration}},}\ }\href {\doibase
  10.1016/j.nuclphysa.2005.03.086} {\bibfield  {journal} {\bibinfo  {journal}
  {Nucl. Phys.}\ }\textbf {\bibinfo {volume} {A757}},\ \bibinfo {pages}
  {184--283} (\bibinfo {year} {2005})},\ \Eprint
  {http://arxiv.org/abs/nucl-ex/0410003} {arXiv:nucl-ex/0410003 [nucl-ex]}
  \BibitemShut {NoStop}%
\bibitem [{\citenamefont {Aad}\ \emph {et~al.}(2010)\citenamefont {Aad} \emph
  {et~al.}}]{Aad:2010bu}%
  \BibitemOpen
  \bibfield  {author} {\bibinfo {author} {\bibfnamefont {Georges}\ \bibnamefont
  {Aad}} \emph {et~al.} (\bibinfo {collaboration} {ATLAS}),\ }\bibfield
  {title} {\enquote {\bibinfo {title} {{Observation of a Centrality-Dependent
  Dijet Asymmetry in Lead-Lead Collisions at $\sqrt{s_{NN}}=2.77$ TeV with the
  ATLAS Detector at the LHC}},}\ }\href {\doibase
  10.1103/PhysRevLett.105.252303} {\bibfield  {journal} {\bibinfo  {journal}
  {Phys. Rev. Lett.}\ }\textbf {\bibinfo {volume} {105}},\ \bibinfo {pages}
  {252303} (\bibinfo {year} {2010})},\ \Eprint {http://arxiv.org/abs/1011.6182}
  {arXiv:1011.6182 [hep-ex]} \BibitemShut {NoStop}%
\bibitem [{\citenamefont {Khachatryan}\ \emph {et~al.}(2010)\citenamefont
  {Khachatryan} \emph {et~al.}}]{Khachatryan:2010gv}%
  \BibitemOpen
  \bibfield  {author} {\bibinfo {author} {\bibfnamefont {Vardan}\ \bibnamefont
  {Khachatryan}} \emph {et~al.} (\bibinfo {collaboration} {CMS}),\ }\bibfield
  {title} {\enquote {\bibinfo {title} {{Observation of Long-Range Near-Side
  Angular Correlations in Proton-Proton Collisions at the LHC}},}\ }\href
  {\doibase 10.1007/JHEP09(2010)091} {\bibfield  {journal} {\bibinfo  {journal}
  {JHEP}\ }\textbf {\bibinfo {volume} {09}},\ \bibinfo {pages} {091} (\bibinfo
  {year} {2010})},\ \Eprint {http://arxiv.org/abs/1009.4122} {arXiv:1009.4122
  [hep-ex]} \BibitemShut {NoStop}%
\bibitem [{\citenamefont {Aad}\ \emph {et~al.}(2013)\citenamefont {Aad} \emph
  {et~al.}}]{Aad:2012gla}%
  \BibitemOpen
  \bibfield  {author} {\bibinfo {author} {\bibfnamefont {Georges}\ \bibnamefont
  {Aad}} \emph {et~al.} (\bibinfo {collaboration} {ATLAS}),\ }\bibfield
  {title} {\enquote {\bibinfo {title} {{Observation of Associated Near-Side and
  Away-Side Long-Range Correlations in $\sqrt{s_{NN}}$=5.02  TeV
  Proton-Lead Collisions with the ATLAS Detector}},}\ }\href {\doibase
  10.1103/PhysRevLett.110.182302} {\bibfield  {journal} {\bibinfo  {journal}
  {Phys. Rev. Lett.}\ }\textbf {\bibinfo {volume} {110}},\ \bibinfo {pages}
  {182302} (\bibinfo {year} {2013})},\ \Eprint {http://arxiv.org/abs/1212.5198}
  {arXiv:1212.5198 [hep-ex]} \BibitemShut {NoStop}%
\bibitem [{\citenamefont {Abelev}\ \emph {et~al.}(2013)\citenamefont {Abelev}
  \emph {et~al.}}]{Abelev:2012ola}%
  \BibitemOpen
  \bibfield  {author} {\bibinfo {author} {\bibfnamefont {Betty}\ \bibnamefont
  {Abelev}} \emph {et~al.} (\bibinfo {collaboration} {ALICE}),\ }\bibfield
  {title} {\enquote {\bibinfo {title} {{Long-range angular correlations on the
  near and away side in $p$-Pb collisions at $\sqrt{s_{NN}}=5.02$ TeV}},}\
  }\href {\doibase 10.1016/j.physletb.2013.01.012} {\bibfield  {journal}
  {\bibinfo  {journal} {Phys. Lett.}\ }\textbf {\bibinfo {volume} {B719}},\
  \bibinfo {pages} {29--41} (\bibinfo {year} {2013})},\ \Eprint
  {http://arxiv.org/abs/1212.2001} {arXiv:1212.2001 [nucl-ex]} \BibitemShut
  {NoStop}%
\bibitem [{\citenamefont {Chatrchyan}\ \emph {et~al.}(2012)\citenamefont
  {Chatrchyan} \emph {et~al.}}]{Chatrchyan:2012lxa}%
  \BibitemOpen
  \bibfield  {author} {\bibinfo {author} {\bibfnamefont {Serguei}\ \bibnamefont
  {Chatrchyan}} \emph {et~al.} (\bibinfo {collaboration} {CMS}),\ }\bibfield
  {title} {\enquote {\bibinfo {title} {{Observation of sequential Upsilon
  suppression in PbPb collisions}},}\ }\href {\doibase
  10.1103/PhysRevLett.109.222301} {\bibfield  {journal} {\bibinfo  {journal}
  {Phys. Rev. Lett.}\ }\textbf {\bibinfo {volume} {109}},\ \bibinfo {pages}
  {222301} (\bibinfo {year} {2012})},\ \Eprint {http://arxiv.org/abs/1208.2826}
  {arXiv:1208.2826 [nucl-ex]} \BibitemShut {NoStop}%
\bibitem [{\citenamefont {Gyulassy}\ and\ \citenamefont
  {McLerran}(2005)}]{Gyulassy:2004zy}%
  \BibitemOpen
  \bibfield  {author} {\bibinfo {author} {\bibfnamefont {Miklos}\ \bibnamefont
  {Gyulassy}}\ and\ \bibinfo {author} {\bibfnamefont {Larry}\ \bibnamefont
  {McLerran}},\ }\bibfield  {title} {\enquote {\bibinfo {title} {{New forms of
  QCD matter discovered at RHIC}},}\ }\bibfield  {booktitle} {\emph {\bibinfo
  {booktitle} {{Quark gluon plasma. New discoveries at RHIC: A case of strongly
  interacting quark gluon plasma. Proceedings, RBRC Workshop, Brookhaven,
  Upton, USA, May 14-15, 2004}}},\ }\href {\doibase
  10.1016/j.nuclphysa.2004.10.034} {\bibfield  {journal} {\bibinfo  {journal}
  {Nucl. Phys.}\ }\textbf {\bibinfo {volume} {A750}},\ \bibinfo {pages}
  {30--63} (\bibinfo {year} {2005})},\ \Eprint
  {http://arxiv.org/abs/nucl-th/0405013} {arXiv:nucl-th/0405013 [nucl-th]}
  \BibitemShut {NoStop}%
\bibitem [{\citenamefont {Teaney}(2003)}]{Teaney:2003kp}%
  \BibitemOpen
  \bibfield  {author} {\bibinfo {author} {\bibfnamefont {Derek}\ \bibnamefont
  {Teaney}},\ }\bibfield  {title} {\enquote {\bibinfo {title} {{The Effects of
  viscosity on spectra, elliptic flow, and HBT radii}},}\ }\href {\doibase
  10.1103/PhysRevC.68.034913} {\bibfield  {journal} {\bibinfo  {journal} {Phys.
  Rev.}\ }\textbf {\bibinfo {volume} {C68}},\ \bibinfo {pages} {034913}
  (\bibinfo {year} {2003})},\ \Eprint {http://arxiv.org/abs/nucl-th/0301099}
  {arXiv:nucl-th/0301099 [nucl-th]} \BibitemShut {NoStop}%
\bibitem [{\citenamefont {Chesler}\ and\ \citenamefont
  {Yaffe}(2011)}]{Chesler:2010bi}%
  \BibitemOpen
  \bibfield  {author} {\bibinfo {author} {\bibfnamefont {Paul~M.}\ \bibnamefont
  {Chesler}}\ and\ \bibinfo {author} {\bibfnamefont {Laurence~G.}\ \bibnamefont
  {Yaffe}},\ }\bibfield  {title} {\enquote {\bibinfo {title} {{Holography and
  colliding gravitational shock waves in asymptotically AdS$_5$ spacetime}},}\
  }\href {\doibase 10.1103/PhysRevLett.106.021601} {\bibfield  {journal}
  {\bibinfo  {journal} {Phys. Rev. Lett.}\ }\textbf {\bibinfo {volume} {106}},\
  \bibinfo {pages} {021601} (\bibinfo {year} {2011})},\ \Eprint
  {http://arxiv.org/abs/1011.3562} {arXiv:1011.3562 [hep-th]} \BibitemShut
  {NoStop}%
\bibitem [{\citenamefont {Song}\ \emph {et~al.}(2011)\citenamefont {Song},
  \citenamefont {Bass}, \citenamefont {Heinz}, \citenamefont {Hirano},\ and\
  \citenamefont {Shen}}]{Song:2010mg}%
  \BibitemOpen
  \bibfield  {author} {\bibinfo {author} {\bibfnamefont {Huichao}\ \bibnamefont
  {Song}}, \bibinfo {author} {\bibfnamefont {Steffen~A.}\ \bibnamefont {Bass}},
  \bibinfo {author} {\bibfnamefont {Ulrich}\ \bibnamefont {Heinz}}, \bibinfo
  {author} {\bibfnamefont {Tetsufumi}\ \bibnamefont {Hirano}}, \ and\ \bibinfo
  {author} {\bibfnamefont {Chun}\ \bibnamefont {Shen}},\ }\bibfield  {title}
  {\enquote {\bibinfo {title} {{200 A GeV Au+Au collisions serve a nearly
  perfect quark-gluon liquid}},}\ }\href {\doibase
  10.1103/PhysRevLett.106.192301, 10.1103/PhysRevLett.109.139904} {\bibfield
  {journal} {\bibinfo  {journal} {Phys. Rev. Lett.}\ }\textbf {\bibinfo
  {volume} {106}},\ \bibinfo {pages} {192301} (\bibinfo {year} {2011})},\
  \bibinfo {note} {[Erratum: Phys. Rev. Lett.109,139904(2012)]},\ \Eprint
  {http://arxiv.org/abs/1011.2783} {arXiv:1011.2783 [nucl-th]} \BibitemShut
  {NoStop}%
\bibitem [{\citenamefont {Gale}\ \emph {et~al.}(2013)\citenamefont {Gale},
  \citenamefont {Jeon}, \citenamefont {Schenke}, \citenamefont {Tribedy},\ and\
  \citenamefont {Venugopalan}}]{Gale:2012rq}%
  \BibitemOpen
  \bibfield  {author} {\bibinfo {author} {\bibfnamefont {Charles}\ \bibnamefont
  {Gale}}, \bibinfo {author} {\bibfnamefont {Sangyong}\ \bibnamefont {Jeon}},
  \bibinfo {author} {\bibfnamefont {Björn}\ \bibnamefont {Schenke}}, \bibinfo
  {author} {\bibfnamefont {Prithwish}\ \bibnamefont {Tribedy}}, \ and\ \bibinfo
  {author} {\bibfnamefont {Raju}\ \bibnamefont {Venugopalan}},\ }\bibfield
  {title} {\enquote {\bibinfo {title} {{Event-by-event anisotropic flow in
  heavy-ion collisions from combined Yang-Mills and viscous fluid dynamics}},}\
  }\href {\doibase 10.1103/PhysRevLett.110.012302} {\bibfield  {journal}
  {\bibinfo  {journal} {Phys. Rev. Lett.}\ }\textbf {\bibinfo {volume} {110}},\
  \bibinfo {pages} {012302} (\bibinfo {year} {2013})},\ \Eprint
  {http://arxiv.org/abs/1209.6330} {arXiv:1209.6330 [nucl-th]} \BibitemShut
  {NoStop}%
\bibitem [{\citenamefont {Bernhard}\ \emph {et~al.}(2016)\citenamefont
  {Bernhard}, \citenamefont {Moreland}, \citenamefont {Bass}, \citenamefont
  {Liu},\ and\ \citenamefont {Heinz}}]{Bernhard:2016tnd}%
  \BibitemOpen
  \bibfield  {author} {\bibinfo {author} {\bibfnamefont {Jonah~E.}\
  \bibnamefont {Bernhard}}, \bibinfo {author} {\bibfnamefont {J.~Scott}\
  \bibnamefont {Moreland}}, \bibinfo {author} {\bibfnamefont {Steffen~A.}\
  \bibnamefont {Bass}}, \bibinfo {author} {\bibfnamefont {Jia}\ \bibnamefont
  {Liu}}, \ and\ \bibinfo {author} {\bibfnamefont {Ulrich}\ \bibnamefont
  {Heinz}},\ }\bibfield  {title} {\enquote {\bibinfo {title} {{Applying
  Bayesian parameter estimation to relativistic heavy-ion collisions:
  simultaneous characterization of the initial state and quark-gluon plasma
  medium}},}\ }\href {\doibase 10.1103/PhysRevC.94.024907} {\bibfield
  {journal} {\bibinfo  {journal} {Phys. Rev.}\ }\textbf {\bibinfo {volume}
  {C94}},\ \bibinfo {pages} {024907} (\bibinfo {year} {2016})},\ \Eprint
  {http://arxiv.org/abs/1605.03954} {arXiv:1605.03954 [nucl-th]} \BibitemShut
  {NoStop}%
\bibitem [{\citenamefont {Alqahtani}\ \emph {et~al.}(2017)\citenamefont
  {Alqahtani}, \citenamefont {Nopoush}, \citenamefont {Ryblewski},\ and\
  \citenamefont {Strickland}}]{Alqahtani:2017tnq}%
  \BibitemOpen
  \bibfield  {author} {\bibinfo {author} {\bibfnamefont {Mubarak}\ \bibnamefont
  {Alqahtani}}, \bibinfo {author} {\bibfnamefont {Mohammad}\ \bibnamefont
  {Nopoush}}, \bibinfo {author} {\bibfnamefont {Radoslaw}\ \bibnamefont
  {Ryblewski}}, \ and\ \bibinfo {author} {\bibfnamefont {Michael}\ \bibnamefont
  {Strickland}},\ }\bibfield  {title} {\enquote {\bibinfo {title} {{Anisotropic
  hydrodynamic modeling of 2.76 TeV Pb-Pb collisions}},}\ }\href@noop {} {\
  (\bibinfo {year} {2017})},\ \Eprint {http://arxiv.org/abs/1705.10191}
  {arXiv:1705.10191 [nucl-th]} \BibitemShut {NoStop}%
\bibitem [{\citenamefont {Morad}\ and\ \citenamefont
  {Horowitz}(2014)}]{Morad:2014xla}%
  \BibitemOpen
  \bibfield  {author} {\bibinfo {author} {\bibfnamefont {R.}~\bibnamefont
  {Morad}}\ and\ \bibinfo {author} {\bibfnamefont {W.~A.}\ \bibnamefont
  {Horowitz}},\ }\bibfield  {title} {\enquote {\bibinfo {title}
  {{Strong-coupling Jet Energy Loss from AdS/CFT}},}\ }\href {\doibase
  10.1007/JHEP11(2014)017} {\bibfield  {journal} {\bibinfo  {journal} {JHEP}\
  }\textbf {\bibinfo {volume} {11}},\ \bibinfo {pages} {017} (\bibinfo {year}
  {2014})},\ \Eprint {http://arxiv.org/abs/1409.7545} {arXiv:1409.7545
  [hep-th]} \BibitemShut {NoStop}%
\bibitem [{\citenamefont {Horowitz}(2015)}]{Horowitz:2015dta}%
  \BibitemOpen
  \bibfield  {author} {\bibinfo {author} {\bibfnamefont {W.~A.}\ \bibnamefont
  {Horowitz}},\ }\bibfield  {title} {\enquote {\bibinfo {title} {{Fluctuating
  heavy quark energy loss in a strongly coupled quark-gluon plasma}},}\ }\href
  {\doibase 10.1103/PhysRevD.91.085019} {\bibfield  {journal} {\bibinfo
  {journal} {Phys. Rev.}\ }\textbf {\bibinfo {volume} {D91}},\ \bibinfo {pages}
  {085019} (\bibinfo {year} {2015})},\ \Eprint
  {http://arxiv.org/abs/1501.04693} {arXiv:1501.04693 [hep-ph]} \BibitemShut
  {NoStop}%
\bibitem [{\citenamefont {Hambrock}\ and\ \citenamefont
  {Horowitz}(2017)}]{Hambrock:2017sno}%
  \BibitemOpen
  \bibfield  {author} {\bibinfo {author} {\bibfnamefont {R.}~\bibnamefont
  {Hambrock}}\ and\ \bibinfo {author} {\bibfnamefont {W.~A.}\ \bibnamefont
  {Horowitz}},\ }\bibfield  {title} {\enquote {\bibinfo {title} {{AdS/CFT
  predictions for azimuthal and momentum correlations of $b\bar{b}$ pairs in
  heavy ion collisions}},}\ }in\ \href
  {https://inspirehep.net/record/1518153/files/arXiv:1703.05845.pdf} {\emph
  {\bibinfo {booktitle} {{8th International Conference on Hard and
  Electromagnetic Probes of High-energy Nuclear Collisions: Hard Probes 2016
  (HP2016) Wuhan, Hubei, China, September 23-27, 2016}}}}\ (\bibinfo {year}
  {2017})\ \Eprint {http://arxiv.org/abs/1703.05845} {arXiv:1703.05845
  [hep-ph]} \BibitemShut {NoStop}%
\bibitem [{\citenamefont {Brewer}\ \emph {et~al.}(2017)\citenamefont {Brewer},
  \citenamefont {Rajagopal}, \citenamefont {Sadofyev},\ and\ \citenamefont
  {van~der Schee}}]{Brewer:2017dwd}%
  \BibitemOpen
  \bibfield  {author} {\bibinfo {author} {\bibfnamefont {Jasmine}\ \bibnamefont
  {Brewer}}, \bibinfo {author} {\bibfnamefont {Krishna}\ \bibnamefont
  {Rajagopal}}, \bibinfo {author} {\bibfnamefont {Andrey}\ \bibnamefont
  {Sadofyev}}, \ and\ \bibinfo {author} {\bibfnamefont {Wilke}\ \bibnamefont
  {van~der Schee}},\ }\bibfield  {title} {\enquote {\bibinfo {title}
  {{Holographic Jet Shapes and their Evolution in Strongly Coupled Plasma}},}\
  }in\ \href {https://inspirehep.net/record/1592399/files/arXiv:1704.05455.pdf}
  {\emph {\bibinfo {booktitle} {{26th International Conference on
  Ultrarelativistic Nucleus-Nucleus Collisions (Quark Matter 2017)
  Chicago,Illinois, USA, February 6-11, 2017}}}}\ (\bibinfo {year} {2017})\
  \Eprint {http://arxiv.org/abs/1704.05455} {arXiv:1704.05455 [nucl-th]}
  \BibitemShut {NoStop}%
\bibitem [{\citenamefont {Weller}\ and\ \citenamefont
  {Romatschke}(2017)}]{Weller:2017tsr}%
  \BibitemOpen
  \bibfield  {author} {\bibinfo {author} {\bibfnamefont {Ryan~D.}\ \bibnamefont
  {Weller}}\ and\ \bibinfo {author} {\bibfnamefont {Paul}\ \bibnamefont
  {Romatschke}},\ }\bibfield  {title} {\enquote {\bibinfo {title} {{One fluid
  to rule them all: viscous hydrodynamic description of event-by-event central
  p+p, p+Pb and Pb+Pb collisions at $\sqrt{s}=5.02$ TeV}},}\ }\href@noop {} {\
  (\bibinfo {year} {2017})},\ \Eprint {http://arxiv.org/abs/1701.07145}
  {arXiv:1701.07145 [nucl-th]} \BibitemShut {NoStop}%
\bibitem [{\citenamefont {Molnar}\ and\ \citenamefont
  {Gyulassy}(2002)}]{Molnar:2001ux}%
  \BibitemOpen
  \bibfield  {author} {\bibinfo {author} {\bibfnamefont {Denes}\ \bibnamefont
  {Molnar}}\ and\ \bibinfo {author} {\bibfnamefont {Miklos}\ \bibnamefont
  {Gyulassy}},\ }\bibfield  {title} {\enquote {\bibinfo {title} {{Saturation of
  elliptic flow and the transport opacity of the gluon plasma at RHIC}},}\
  }\href {\doibase 10.1016/S0375-9474(01)01224-6,
  10.1016/S0375-9474(02)00859-X} {\bibfield  {journal} {\bibinfo  {journal}
  {Nucl. Phys.}\ }\textbf {\bibinfo {volume} {A697}},\ \bibinfo {pages}
  {495--520} (\bibinfo {year} {2002})},\ \bibinfo {note} {[Erratum: Nucl.
  Phys.A703,893(2002)]},\ \Eprint {http://arxiv.org/abs/nucl-th/0104073}
  {arXiv:nucl-th/0104073 [nucl-th]} \BibitemShut {NoStop}%
\bibitem [{\citenamefont {Lin}\ and\ \citenamefont {Ko}(2002)}]{Lin:2001zk}%
  \BibitemOpen
  \bibfield  {author} {\bibinfo {author} {\bibfnamefont {Zi-wei}\ \bibnamefont
  {Lin}}\ and\ \bibinfo {author} {\bibfnamefont {C.~M.}\ \bibnamefont {Ko}},\
  }\bibfield  {title} {\enquote {\bibinfo {title} {{Partonic effects on the
  elliptic flow at RHIC}},}\ }\href {\doibase 10.1103/PhysRevC.65.034904}
  {\bibfield  {journal} {\bibinfo  {journal} {Phys. Rev.}\ }\textbf {\bibinfo
  {volume} {C65}},\ \bibinfo {pages} {034904} (\bibinfo {year} {2002})},\
  \Eprint {http://arxiv.org/abs/nucl-th/0108039} {arXiv:nucl-th/0108039
  [nucl-th]} \BibitemShut {NoStop}%
\bibitem [{\citenamefont {Bzdak}\ and\ \citenamefont
  {Ma}(2014)}]{Bzdak:2014dia}%
  \BibitemOpen
  \bibfield  {author} {\bibinfo {author} {\bibfnamefont {Adam}\ \bibnamefont
  {Bzdak}}\ and\ \bibinfo {author} {\bibfnamefont {Guo-Liang}\ \bibnamefont
  {Ma}},\ }\bibfield  {title} {\enquote {\bibinfo {title} {{Elliptic and
  triangular flow in $p$+Pb and peripheral Pb+Pb collisions from parton
  scatterings}},}\ }\href {\doibase 10.1103/PhysRevLett.113.252301} {\bibfield
  {journal} {\bibinfo  {journal} {Phys. Rev. Lett.}\ }\textbf {\bibinfo
  {volume} {113}},\ \bibinfo {pages} {252301} (\bibinfo {year} {2014})},\
  \Eprint {http://arxiv.org/abs/1406.2804} {arXiv:1406.2804 [hep-ph]}
  \BibitemShut {NoStop}%
\bibitem [{\citenamefont {Orjuela~Koop}\ \emph {et~al.}(2015)\citenamefont
  {Orjuela~Koop}, \citenamefont {Adare}, \citenamefont {McGlinchey},\ and\
  \citenamefont {Nagle}}]{Koop:2015wea}%
  \BibitemOpen
  \bibfield  {author} {\bibinfo {author} {\bibfnamefont {J.~D.}\ \bibnamefont
  {Orjuela~Koop}}, \bibinfo {author} {\bibfnamefont {A.}~\bibnamefont {Adare}},
  \bibinfo {author} {\bibfnamefont {D.}~\bibnamefont {McGlinchey}}, \ and\
  \bibinfo {author} {\bibfnamefont {J.~L.}\ \bibnamefont {Nagle}},\ }\bibfield
  {title} {\enquote {\bibinfo {title} {{Azimuthal anisotropy relative to the
  participant plane from a multiphase transport model in central p + Au , d +
  Au , and $^{3}$He + Au collisions at $\sqrt{s_{NN}}=200$ GeV}},}\ }\href
  {\doibase 10.1103/PhysRevC.92.054903} {\bibfield  {journal} {\bibinfo
  {journal} {Phys. Rev.}\ }\textbf {\bibinfo {volume} {C92}},\ \bibinfo {pages}
  {054903} (\bibinfo {year} {2015})},\ \Eprint
  {http://arxiv.org/abs/1501.06880} {arXiv:1501.06880 [nucl-ex]} \BibitemShut
  {NoStop}%
\bibitem [{\citenamefont {He}\ \emph {et~al.}(2016)\citenamefont {He},
  \citenamefont {Edmonds}, \citenamefont {Lin}, \citenamefont {Liu},
  \citenamefont {Molnar},\ and\ \citenamefont {Wang}}]{He:2015hfa}%
  \BibitemOpen
  \bibfield  {author} {\bibinfo {author} {\bibfnamefont {Liang}\ \bibnamefont
  {He}}, \bibinfo {author} {\bibfnamefont {Terrence}\ \bibnamefont {Edmonds}},
  \bibinfo {author} {\bibfnamefont {Zi-Wei}\ \bibnamefont {Lin}}, \bibinfo
  {author} {\bibfnamefont {Feng}\ \bibnamefont {Liu}}, \bibinfo {author}
  {\bibfnamefont {Denes}\ \bibnamefont {Molnar}}, \ and\ \bibinfo {author}
  {\bibfnamefont {Fuqiang}\ \bibnamefont {Wang}},\ }\bibfield  {title}
  {\enquote {\bibinfo {title} {{Anisotropic parton escape is the dominant
  source of azimuthal anisotropy in transport models}},}\ }\href {\doibase
  10.1016/j.physletb.2015.12.051} {\bibfield  {journal} {\bibinfo  {journal}
  {Phys. Lett.}\ }\textbf {\bibinfo {volume} {B753}},\ \bibinfo {pages}
  {506--510} (\bibinfo {year} {2016})},\ \Eprint
  {http://arxiv.org/abs/1502.05572} {arXiv:1502.05572 [nucl-th]} \BibitemShut
  {NoStop}%
\bibitem [{\citenamefont {Lin}\ \emph {et~al.}(2016)\citenamefont {Lin},
  \citenamefont {He}, \citenamefont {Edmonds}, \citenamefont {Liu},
  \citenamefont {Molnar},\ and\ \citenamefont {Wang}}]{Lin:2015ucn}%
  \BibitemOpen
  \bibfield  {author} {\bibinfo {author} {\bibfnamefont {Zi-Wei}\ \bibnamefont
  {Lin}}, \bibinfo {author} {\bibfnamefont {Liang}\ \bibnamefont {He}},
  \bibinfo {author} {\bibfnamefont {Terrence}\ \bibnamefont {Edmonds}},
  \bibinfo {author} {\bibfnamefont {Feng}\ \bibnamefont {Liu}}, \bibinfo
  {author} {\bibfnamefont {Denes}\ \bibnamefont {Molnar}}, \ and\ \bibinfo
  {author} {\bibfnamefont {Fuqiang}\ \bibnamefont {Wang}},\ }\bibfield  {title}
  {\enquote {\bibinfo {title} {{Elliptic Anisotropy $v_2$ May Be Dominated by
  Particle Escape instead of Hydrodynamic Flow}},}\ }\bibfield  {booktitle}
  {\emph {\bibinfo {booktitle} {{Proceedings, 25th International Conference on
  Ultra-Relativistic Nucleus-Nucleus Collisions (Quark Matter 2015): Kobe,
  Japan, September 27-October 3, 2015}}},\ }\href {\doibase
  10.1016/j.nuclphysa.2016.01.017} {\bibfield  {journal} {\bibinfo  {journal}
  {Nucl. Phys.}\ }\textbf {\bibinfo {volume} {A956}},\ \bibinfo {pages}
  {316--319} (\bibinfo {year} {2016})},\ \Eprint
  {http://arxiv.org/abs/1512.06465} {arXiv:1512.06465 [nucl-th]} \BibitemShut
  {NoStop}%
\bibitem [{\citenamefont {Gyulassy}\ \emph {et~al.}(2002)\citenamefont
  {Gyulassy}, \citenamefont {Levai},\ and\ \citenamefont
  {Vitev}}]{Gyulassy:2001nm}%
  \BibitemOpen
  \bibfield  {author} {\bibinfo {author} {\bibfnamefont {M.}~\bibnamefont
  {Gyulassy}}, \bibinfo {author} {\bibfnamefont {P.}~\bibnamefont {Levai}}, \
  and\ \bibinfo {author} {\bibfnamefont {I.}~\bibnamefont {Vitev}},\ }\bibfield
   {title} {\enquote {\bibinfo {title} {{Jet tomography of Au+Au reactions
  including multigluon fluctuations}},}\ }\href {\doibase
  10.1016/S0370-2693(02)01990-1} {\bibfield  {journal} {\bibinfo  {journal}
  {Phys. Lett.}\ }\textbf {\bibinfo {volume} {B538}},\ \bibinfo {pages}
  {282--288} (\bibinfo {year} {2002})},\ \Eprint
  {http://arxiv.org/abs/nucl-th/0112071} {arXiv:nucl-th/0112071 [nucl-th]}
  \BibitemShut {NoStop}%
\bibitem [{\citenamefont {Vitev}\ and\ \citenamefont
  {Gyulassy}(2002)}]{Vitev:2002pf}%
  \BibitemOpen
  \bibfield  {author} {\bibinfo {author} {\bibfnamefont {Ivan}\ \bibnamefont
  {Vitev}}\ and\ \bibinfo {author} {\bibfnamefont {Miklos}\ \bibnamefont
  {Gyulassy}},\ }\bibfield  {title} {\enquote {\bibinfo {title} {{High $p_{T}$
  tomography of $d$ + Au and Au+Au at SPS, RHIC, and LHC}},}\ }\href {\doibase
  10.1103/PhysRevLett.89.252301} {\bibfield  {journal} {\bibinfo  {journal}
  {Phys. Rev. Lett.}\ }\textbf {\bibinfo {volume} {89}},\ \bibinfo {pages}
  {252301} (\bibinfo {year} {2002})},\ \Eprint
  {http://arxiv.org/abs/hep-ph/0209161} {arXiv:hep-ph/0209161 [hep-ph]}
  \BibitemShut {NoStop}%
\bibitem [{\citenamefont {Wang}\ and\ \citenamefont
  {Wang}(2002)}]{Wang:2002ri}%
  \BibitemOpen
  \bibfield  {author} {\bibinfo {author} {\bibfnamefont {Enke}\ \bibnamefont
  {Wang}}\ and\ \bibinfo {author} {\bibfnamefont {Xin-Nian}\ \bibnamefont
  {Wang}},\ }\bibfield  {title} {\enquote {\bibinfo {title} {{Jet tomography of
  dense and nuclear matter}},}\ }\href {\doibase 10.1103/PhysRevLett.89.162301}
  {\bibfield  {journal} {\bibinfo  {journal} {Phys. Rev. Lett.}\ }\textbf
  {\bibinfo {volume} {89}},\ \bibinfo {pages} {162301} (\bibinfo {year}
  {2002})},\ \Eprint {http://arxiv.org/abs/hep-ph/0202105}
  {arXiv:hep-ph/0202105 [hep-ph]} \BibitemShut {NoStop}%
\bibitem [{\citenamefont {Majumder}\ \emph
  {et~al.}(2007{\natexlab{a}})\citenamefont {Majumder}, \citenamefont {Wang},\
  and\ \citenamefont {Wang}}]{Majumder:2004pt}%
  \BibitemOpen
  \bibfield  {author} {\bibinfo {author} {\bibfnamefont {A.}~\bibnamefont
  {Majumder}}, \bibinfo {author} {\bibfnamefont {Enke}\ \bibnamefont {Wang}}, \
  and\ \bibinfo {author} {\bibfnamefont {Xin-Nian}\ \bibnamefont {Wang}},\
  }\bibfield  {title} {\enquote {\bibinfo {title} {{Modified dihadron
  fragmentation functions in hot and nuclear matter}},}\ }\href {\doibase
  10.1103/PhysRevLett.99.152301} {\bibfield  {journal} {\bibinfo  {journal}
  {Phys. Rev. Lett.}\ }\textbf {\bibinfo {volume} {99}},\ \bibinfo {pages}
  {152301} (\bibinfo {year} {2007}{\natexlab{a}})},\ \Eprint
  {http://arxiv.org/abs/nucl-th/0412061} {arXiv:nucl-th/0412061 [nucl-th]}
  \BibitemShut {NoStop}%
\bibitem [{\citenamefont {Dainese}\ \emph {et~al.}(2005)\citenamefont
  {Dainese}, \citenamefont {Loizides},\ and\ \citenamefont
  {Paic}}]{Dainese:2004te}%
  \BibitemOpen
  \bibfield  {author} {\bibinfo {author} {\bibfnamefont {A.}~\bibnamefont
  {Dainese}}, \bibinfo {author} {\bibfnamefont {C.}~\bibnamefont {Loizides}}, \
  and\ \bibinfo {author} {\bibfnamefont {G.}~\bibnamefont {Paic}},\ }\bibfield
  {title} {\enquote {\bibinfo {title} {{Leading-particle suppression in high
  energy nucleus-nucleus collisions}},}\ }\href {\doibase
  10.1140/epjc/s2004-02077-x} {\bibfield  {journal} {\bibinfo  {journal} {Eur.
  Phys. J.}\ }\textbf {\bibinfo {volume} {C38}},\ \bibinfo {pages} {461--474}
  (\bibinfo {year} {2005})},\ \Eprint {http://arxiv.org/abs/hep-ph/0406201}
  {arXiv:hep-ph/0406201 [hep-ph]} \BibitemShut {NoStop}%
\bibitem [{\citenamefont {Armesto}\ \emph {et~al.}(2006)\citenamefont
  {Armesto}, \citenamefont {Cacciari}, \citenamefont {Dainese}, \citenamefont
  {Salgado},\ and\ \citenamefont {Wiedemann}}]{Armesto:2005mz}%
  \BibitemOpen
  \bibfield  {author} {\bibinfo {author} {\bibfnamefont {Nestor}\ \bibnamefont
  {Armesto}}, \bibinfo {author} {\bibfnamefont {Matteo}\ \bibnamefont
  {Cacciari}}, \bibinfo {author} {\bibfnamefont {Andrea}\ \bibnamefont
  {Dainese}}, \bibinfo {author} {\bibfnamefont {Carlos~A.}\ \bibnamefont
  {Salgado}}, \ and\ \bibinfo {author} {\bibfnamefont {Urs~Achim}\ \bibnamefont
  {Wiedemann}},\ }\bibfield  {title} {\enquote {\bibinfo {title} {{How
  sensitive are high-p(T) electron spectra at RHIC to heavy quark energy
  loss?}}}\ }\href {\doibase 10.1016/j.physletb.2005.12.073} {\bibfield
  {journal} {\bibinfo  {journal} {Phys. Lett.}\ }\textbf {\bibinfo {volume}
  {B637}},\ \bibinfo {pages} {362--366} (\bibinfo {year} {2006})},\ \Eprint
  {http://arxiv.org/abs/hep-ph/0511257} {arXiv:hep-ph/0511257 [hep-ph]}
  \BibitemShut {NoStop}%
\bibitem [{\citenamefont {Wicks}\ \emph {et~al.}(2007)\citenamefont {Wicks},
  \citenamefont {Horowitz}, \citenamefont {Djordjevic},\ and\ \citenamefont
  {Gyulassy}}]{Wicks:2005gt}%
  \BibitemOpen
  \bibfield  {author} {\bibinfo {author} {\bibfnamefont {Simon}\ \bibnamefont
  {Wicks}}, \bibinfo {author} {\bibfnamefont {William}\ \bibnamefont
  {Horowitz}}, \bibinfo {author} {\bibfnamefont {Magdalena}\ \bibnamefont
  {Djordjevic}}, \ and\ \bibinfo {author} {\bibfnamefont {Miklos}\ \bibnamefont
  {Gyulassy}},\ }\bibfield  {title} {\enquote {\bibinfo {title} {{Elastic,
  inelastic, and path length fluctuations in jet tomography}},}\ }\href
  {\doibase 10.1016/j.nuclphysa.2006.12.048} {\bibfield  {journal} {\bibinfo
  {journal} {Nucl. Phys.}\ }\textbf {\bibinfo {volume} {A784}},\ \bibinfo
  {pages} {426--442} (\bibinfo {year} {2007})},\ \Eprint
  {http://arxiv.org/abs/nucl-th/0512076} {arXiv:nucl-th/0512076 [nucl-th]}
  \BibitemShut {NoStop}%
\bibitem [{\citenamefont {Majumder}\ \emph
  {et~al.}(2007{\natexlab{b}})\citenamefont {Majumder}, \citenamefont
  {Nonaka},\ and\ \citenamefont {Bass}}]{Majumder:2007ae}%
  \BibitemOpen
  \bibfield  {author} {\bibinfo {author} {\bibfnamefont {A.}~\bibnamefont
  {Majumder}}, \bibinfo {author} {\bibfnamefont {C.}~\bibnamefont {Nonaka}}, \
  and\ \bibinfo {author} {\bibfnamefont {S.~A.}\ \bibnamefont {Bass}},\
  }\bibfield  {title} {\enquote {\bibinfo {title} {{Jet modification in three
  dimensional fluid dynamics at next-to-leading twist}},}\ }\href {\doibase
  10.1103/PhysRevC.76.041902} {\bibfield  {journal} {\bibinfo  {journal} {Phys.
  Rev.}\ }\textbf {\bibinfo {volume} {C76}},\ \bibinfo {pages} {041902}
  (\bibinfo {year} {2007}{\natexlab{b}})},\ \Eprint
  {http://arxiv.org/abs/nucl-th/0703019} {arXiv:nucl-th/0703019 [nucl-th]}
  \BibitemShut {NoStop}%
\bibitem [{\citenamefont {Zhang}\ \emph {et~al.}(2009)\citenamefont {Zhang},
  \citenamefont {Owens}, \citenamefont {Wang},\ and\ \citenamefont
  {Wang}}]{Zhang:2009rn}%
  \BibitemOpen
  \bibfield  {author} {\bibinfo {author} {\bibfnamefont {Hanzhong}\
  \bibnamefont {Zhang}}, \bibinfo {author} {\bibfnamefont {J.~F.}\ \bibnamefont
  {Owens}}, \bibinfo {author} {\bibfnamefont {Enke}\ \bibnamefont {Wang}}, \
  and\ \bibinfo {author} {\bibfnamefont {Xin-Nian}\ \bibnamefont {Wang}},\
  }\bibfield  {title} {\enquote {\bibinfo {title} {{Tomography of high-energy
  nuclear collisions with photon-hadron correlations}},}\ }\href {\doibase
  10.1103/PhysRevLett.103.032302} {\bibfield  {journal} {\bibinfo  {journal}
  {Phys. Rev. Lett.}\ }\textbf {\bibinfo {volume} {103}},\ \bibinfo {pages}
  {032302} (\bibinfo {year} {2009})},\ \Eprint {http://arxiv.org/abs/0902.4000}
  {arXiv:0902.4000 [nucl-th]} \BibitemShut {NoStop}%
\bibitem [{\citenamefont {Vitev}\ and\ \citenamefont
  {Zhang}(2010)}]{Vitev:2009rd}%
  \BibitemOpen
  \bibfield  {author} {\bibinfo {author} {\bibfnamefont {Ivan}\ \bibnamefont
  {Vitev}}\ and\ \bibinfo {author} {\bibfnamefont {Ben-Wei}\ \bibnamefont
  {Zhang}},\ }\bibfield  {title} {\enquote {\bibinfo {title} {{Jet tomography
  of high-energy nucleus-nucleus collisions at next-to-leading order}},}\
  }\href {\doibase 10.1103/PhysRevLett.104.132001} {\bibfield  {journal}
  {\bibinfo  {journal} {Phys. Rev. Lett.}\ }\textbf {\bibinfo {volume} {104}},\
  \bibinfo {pages} {132001} (\bibinfo {year} {2010})},\ \Eprint
  {http://arxiv.org/abs/0910.1090} {arXiv:0910.1090 [hep-ph]} \BibitemShut
  {NoStop}%
\bibitem [{\citenamefont {Schenke}\ \emph {et~al.}(2009)\citenamefont
  {Schenke}, \citenamefont {Gale},\ and\ \citenamefont
  {Jeon}}]{Schenke:2009gb}%
  \BibitemOpen
  \bibfield  {author} {\bibinfo {author} {\bibfnamefont {Bjoern}\ \bibnamefont
  {Schenke}}, \bibinfo {author} {\bibfnamefont {Charles}\ \bibnamefont {Gale}},
  \ and\ \bibinfo {author} {\bibfnamefont {Sangyong}\ \bibnamefont {Jeon}},\
  }\bibfield  {title} {\enquote {\bibinfo {title} {{MARTINI: An Event generator
  for relativistic heavy-ion collisions}},}\ }\href {\doibase
  10.1103/PhysRevC.80.054913} {\bibfield  {journal} {\bibinfo  {journal} {Phys.
  Rev.}\ }\textbf {\bibinfo {volume} {C80}},\ \bibinfo {pages} {054913}
  (\bibinfo {year} {2009})},\ \Eprint {http://arxiv.org/abs/0909.2037}
  {arXiv:0909.2037 [hep-ph]} \BibitemShut {NoStop}%
\bibitem [{\citenamefont {Young}\ \emph {et~al.}(2011)\citenamefont {Young},
  \citenamefont {Schenke}, \citenamefont {Jeon},\ and\ \citenamefont
  {Gale}}]{Young:2011qx}%
  \BibitemOpen
  \bibfield  {author} {\bibinfo {author} {\bibfnamefont {Clint}\ \bibnamefont
  {Young}}, \bibinfo {author} {\bibfnamefont {Bjorn}\ \bibnamefont {Schenke}},
  \bibinfo {author} {\bibfnamefont {Sangyong}\ \bibnamefont {Jeon}}, \ and\
  \bibinfo {author} {\bibfnamefont {Charles}\ \bibnamefont {Gale}},\ }\bibfield
   {title} {\enquote {\bibinfo {title} {{Dijet asymmetry at the energies
  available at the CERN Large Hadron Collider}},}\ }\href {\doibase
  10.1103/PhysRevC.84.024907} {\bibfield  {journal} {\bibinfo  {journal} {Phys.
  Rev.}\ }\textbf {\bibinfo {volume} {C84}},\ \bibinfo {pages} {024907}
  (\bibinfo {year} {2011})},\ \Eprint {http://arxiv.org/abs/1103.5769}
  {arXiv:1103.5769 [nucl-th]} \BibitemShut {NoStop}%
\bibitem [{\citenamefont {Majumder}\ and\ \citenamefont
  {Shen}(2012)}]{Majumder:2011uk}%
  \BibitemOpen
  \bibfield  {author} {\bibinfo {author} {\bibfnamefont {A.}~\bibnamefont
  {Majumder}}\ and\ \bibinfo {author} {\bibfnamefont {C.}~\bibnamefont
  {Shen}},\ }\bibfield  {title} {\enquote {\bibinfo {title} {{Suppression of
  the High $p_T$ Charged Hadron $R_{AA}$ at the LHC}},}\ }\href {\doibase
  10.1103/PhysRevLett.109.202301} {\bibfield  {journal} {\bibinfo  {journal}
  {Phys. Rev. Lett.}\ }\textbf {\bibinfo {volume} {109}},\ \bibinfo {pages}
  {202301} (\bibinfo {year} {2012})},\ \Eprint {http://arxiv.org/abs/1103.0809}
  {arXiv:1103.0809 [hep-ph]} \BibitemShut {NoStop}%
\bibitem [{\citenamefont {Horowitz}\ and\ \citenamefont
  {Gyulassy}(2011)}]{Horowitz:2011gd}%
  \BibitemOpen
  \bibfield  {author} {\bibinfo {author} {\bibfnamefont {W.~A.}\ \bibnamefont
  {Horowitz}}\ and\ \bibinfo {author} {\bibfnamefont {Miklos}\ \bibnamefont
  {Gyulassy}},\ }\bibfield  {title} {\enquote {\bibinfo {title} {{The
  Surprising Transparency of the sQGP at LHC}},}\ }\href {\doibase
  10.1016/j.nuclphysa.2011.09.018} {\bibfield  {journal} {\bibinfo  {journal}
  {Nucl. Phys.}\ }\textbf {\bibinfo {volume} {A872}},\ \bibinfo {pages}
  {265--285} (\bibinfo {year} {2011})},\ \Eprint
  {http://arxiv.org/abs/1104.4958} {arXiv:1104.4958 [hep-ph]} \BibitemShut
  {NoStop}%
\bibitem [{\citenamefont {Buzzatti}\ and\ \citenamefont
  {Gyulassy}(2012)}]{Buzzatti:2011vt}%
  \BibitemOpen
  \bibfield  {author} {\bibinfo {author} {\bibfnamefont {Alessandro}\
  \bibnamefont {Buzzatti}}\ and\ \bibinfo {author} {\bibfnamefont {Miklos}\
  \bibnamefont {Gyulassy}},\ }\bibfield  {title} {\enquote {\bibinfo {title}
  {{Jet Flavor Tomography of Quark Gluon Plasmas at RHIC and LHC}},}\ }\href
  {\doibase 10.1103/PhysRevLett.108.022301} {\bibfield  {journal} {\bibinfo
  {journal} {Phys. Rev. Lett.}\ }\textbf {\bibinfo {volume} {108}},\ \bibinfo
  {pages} {022301} (\bibinfo {year} {2012})},\ \Eprint
  {http://arxiv.org/abs/1106.3061} {arXiv:1106.3061 [hep-ph]} \BibitemShut
  {NoStop}%
\bibitem [{\citenamefont {Horowitz}(2013)}]{Horowitz:2012cf}%
  \BibitemOpen
  \bibfield  {author} {\bibinfo {author} {\bibfnamefont {W.~A.}\ \bibnamefont
  {Horowitz}},\ }\bibfield  {title} {\enquote {\bibinfo {title} {{Heavy Quark
  Production and Energy Loss}},}\ }\bibfield  {booktitle} {\emph {\bibinfo
  {booktitle} {{Proceedings, 23rd International Conference on Ultrarelativistic
  Nucleus-Nucleus Collisions : Quark Matter 2012 (QM 2012): Washington, DC,
  USA, August 13-18, 2012}}},\ }\href {\doibase
  10.1016/j.nuclphysa.2013.01.061} {\bibfield  {journal} {\bibinfo  {journal}
  {Nucl. Phys.}\ }\textbf {\bibinfo {volume} {A904-905}},\ \bibinfo {pages}
  {186c--193c} (\bibinfo {year} {2013})},\ \Eprint
  {http://arxiv.org/abs/1210.8330} {arXiv:1210.8330 [nucl-th]} \BibitemShut
  {NoStop}%
\bibitem [{\citenamefont {Djordjevic}\ and\ \citenamefont
  {Djordjevic}(2014)}]{Djordjevic:2013xoa}%
  \BibitemOpen
  \bibfield  {author} {\bibinfo {author} {\bibfnamefont {Magdalena}\
  \bibnamefont {Djordjevic}}\ and\ \bibinfo {author} {\bibfnamefont {Marko}\
  \bibnamefont {Djordjevic}},\ }\bibfield  {title} {\enquote {\bibinfo {title}
  {{LHC jet suppression of light and heavy flavor observables}},}\ }\href
  {\doibase 10.1016/j.physletb.2014.05.053} {\bibfield  {journal} {\bibinfo
  {journal} {Phys. Lett.}\ }\textbf {\bibinfo {volume} {B734}},\ \bibinfo
  {pages} {286--289} (\bibinfo {year} {2014})},\ \Eprint
  {http://arxiv.org/abs/1307.4098} {arXiv:1307.4098 [hep-ph]} \BibitemShut
  {NoStop}%
\bibitem [{\citenamefont {Liu}\ \emph {et~al.}(2006)\citenamefont {Liu},
  \citenamefont {Rajagopal},\ and\ \citenamefont {Wiedemann}}]{Liu:2006ug}%
  \BibitemOpen
  \bibfield  {author} {\bibinfo {author} {\bibfnamefont {Hong}\ \bibnamefont
  {Liu}}, \bibinfo {author} {\bibfnamefont {Krishna}\ \bibnamefont
  {Rajagopal}}, \ and\ \bibinfo {author} {\bibfnamefont {Urs~Achim}\
  \bibnamefont {Wiedemann}},\ }\bibfield  {title} {\enquote {\bibinfo {title}
  {{Calculating the jet quenching parameter from AdS/CFT}},}\ }\href {\doibase
  10.1103/PhysRevLett.97.182301} {\bibfield  {journal} {\bibinfo  {journal}
  {Phys. Rev. Lett.}\ }\textbf {\bibinfo {volume} {97}},\ \bibinfo {pages}
  {182301} (\bibinfo {year} {2006})},\ \Eprint
  {http://arxiv.org/abs/hep-ph/0605178} {arXiv:hep-ph/0605178 [hep-ph]}
  \BibitemShut {NoStop}%
\bibitem [{\citenamefont {Casalderrey-Solana}\ \emph
  {et~al.}(2014)\citenamefont {Casalderrey-Solana}, \citenamefont {Gulhan},
  \citenamefont {Milhano}, \citenamefont {Pablos},\ and\ \citenamefont
  {Rajagopal}}]{Casalderrey-Solana:2014bpa}%
  \BibitemOpen
  \bibfield  {author} {\bibinfo {author} {\bibfnamefont {Jorge}\ \bibnamefont
  {Casalderrey-Solana}}, \bibinfo {author} {\bibfnamefont {Doga~Can}\
  \bibnamefont {Gulhan}}, \bibinfo {author} {\bibfnamefont {José~Guilherme}\
  \bibnamefont {Milhano}}, \bibinfo {author} {\bibfnamefont {Daniel}\
  \bibnamefont {Pablos}}, \ and\ \bibinfo {author} {\bibfnamefont {Krishna}\
  \bibnamefont {Rajagopal}},\ }\bibfield  {title} {\enquote {\bibinfo {title}
  {{A Hybrid Strong/Weak Coupling Approach to Jet Quenching}},}\ }\href
  {\doibase 10.1007/JHEP09(2015)175, 10.1007/JHEP10(2014)019} {\bibfield
  {journal} {\bibinfo  {journal} {JHEP}\ }\textbf {\bibinfo {volume} {10}},\
  \bibinfo {pages} {019} (\bibinfo {year} {2014})},\ \bibinfo {note} {[Erratum:
  JHEP09,175(2015)]},\ \Eprint {http://arxiv.org/abs/1405.3864}
  {arXiv:1405.3864 [hep-ph]} \BibitemShut {NoStop}%
\bibitem [{\citenamefont {Casalderrey-Solana}\ \emph
  {et~al.}(2016)\citenamefont {Casalderrey-Solana}, \citenamefont {Gulhan},
  \citenamefont {Milhano}, \citenamefont {Pablos},\ and\ \citenamefont
  {Rajagopal}}]{Casalderrey-Solana:2015vaa}%
  \BibitemOpen
  \bibfield  {author} {\bibinfo {author} {\bibfnamefont {Jorge}\ \bibnamefont
  {Casalderrey-Solana}}, \bibinfo {author} {\bibfnamefont {Doga~Can}\
  \bibnamefont {Gulhan}}, \bibinfo {author} {\bibfnamefont {José~Guilherme}\
  \bibnamefont {Milhano}}, \bibinfo {author} {\bibfnamefont {Daniel}\
  \bibnamefont {Pablos}}, \ and\ \bibinfo {author} {\bibfnamefont {Krishna}\
  \bibnamefont {Rajagopal}},\ }\bibfield  {title} {\enquote {\bibinfo {title}
  {{Predictions for Boson-Jet Observables and Fragmentation Function Ratios
  from a Hybrid Strong/Weak Coupling Model for Jet Quenching}},}\ }\href
  {\doibase 10.1007/JHEP03(2016)053} {\bibfield  {journal} {\bibinfo  {journal}
  {JHEP}\ }\textbf {\bibinfo {volume} {03}},\ \bibinfo {pages} {053} (\bibinfo
  {year} {2016})},\ \Eprint {http://arxiv.org/abs/1508.00815} {arXiv:1508.00815
  [hep-ph]} \BibitemShut {NoStop}%
\bibitem [{\citenamefont {Casalderrey-Solana}\ \emph
  {et~al.}(2017)\citenamefont {Casalderrey-Solana}, \citenamefont {Gulhan},
  \citenamefont {Milhano}, \citenamefont {Pablos},\ and\ \citenamefont
  {Rajagopal}}]{Casalderrey-Solana:2016jvj}%
  \BibitemOpen
  \bibfield  {author} {\bibinfo {author} {\bibfnamefont {Jorge}\ \bibnamefont
  {Casalderrey-Solana}}, \bibinfo {author} {\bibfnamefont {Doga}\ \bibnamefont
  {Gulhan}}, \bibinfo {author} {\bibfnamefont {Guilherme}\ \bibnamefont
  {Milhano}}, \bibinfo {author} {\bibfnamefont {Daniel}\ \bibnamefont
  {Pablos}}, \ and\ \bibinfo {author} {\bibfnamefont {Krishna}\ \bibnamefont
  {Rajagopal}},\ }\bibfield  {title} {\enquote {\bibinfo {title} {{Angular
  Structure of Jet Quenching Within a Hybrid Strong/Weak Coupling Model}},}\
  }\href {\doibase 10.1007/JHEP03(2017)135} {\bibfield  {journal} {\bibinfo
  {journal} {JHEP}\ }\textbf {\bibinfo {volume} {03}},\ \bibinfo {pages} {135}
  (\bibinfo {year} {2017})},\ \Eprint {http://arxiv.org/abs/1609.05842}
  {arXiv:1609.05842 [hep-ph]} \BibitemShut {NoStop}%
\bibitem [{\citenamefont {Patrignani}\ \emph {et~al.}(2016)\citenamefont
  {Patrignani} \emph {et~al.}}]{Olive:2016xmw}%
  \BibitemOpen
  \bibfield  {author} {\bibinfo {author} {\bibfnamefont {C.}~\bibnamefont
  {Patrignani}} \emph {et~al.} (\bibinfo {collaboration} {Particle Data
  Group}),\ }\bibfield  {title} {\enquote {\bibinfo {title} {{Review of
  Particle Physics}},}\ }\href {\doibase 10.1088/1674-1137/40/10/100001}
  {\bibfield  {journal} {\bibinfo  {journal} {Chin. Phys.}\ }\textbf {\bibinfo
  {volume} {C40}},\ \bibinfo {pages} {100001} (\bibinfo {year}
  {2016})}\BibitemShut {NoStop}%
\bibitem [{\citenamefont {Matsui}\ and\ \citenamefont
  {Satz}(1986)}]{Matsui:1986dk}%
  \BibitemOpen
  \bibfield  {author} {\bibinfo {author} {\bibfnamefont {T.}~\bibnamefont
  {Matsui}}\ and\ \bibinfo {author} {\bibfnamefont {H.}~\bibnamefont {Satz}},\
  }\bibfield  {title} {\enquote {\bibinfo {title} {{$J/\psi$ Suppression by
  Quark-Gluon Plasma Formation}},}\ }\href {\doibase
  10.1016/0370-2693(86)91404-8} {\bibfield  {journal} {\bibinfo  {journal}
  {Phys. Lett.}\ }\textbf {\bibinfo {volume} {B178}},\ \bibinfo {pages}
  {416--422} (\bibinfo {year} {1986})}\BibitemShut {NoStop}%
\bibitem [{\citenamefont {Karsch}\ \emph {et~al.}(1988)\citenamefont {Karsch},
  \citenamefont {Mehr},\ and\ \citenamefont {Satz}}]{Karsch:1987pv}%
  \BibitemOpen
  \bibfield  {author} {\bibinfo {author} {\bibfnamefont {F.}~\bibnamefont
  {Karsch}}, \bibinfo {author} {\bibfnamefont {M.~T.}\ \bibnamefont {Mehr}}, \
  and\ \bibinfo {author} {\bibfnamefont {H.}~\bibnamefont {Satz}},\ }\bibfield
  {title} {\enquote {\bibinfo {title} {{Color Screening and Deconfinement for
  Bound States of Heavy Quarks}},}\ }\href {\doibase 10.1007/BF01549722}
  {\bibfield  {journal} {\bibinfo  {journal} {Z. Phys.}\ }\textbf {\bibinfo
  {volume} {C37}},\ \bibinfo {pages} {617} (\bibinfo {year}
  {1988})}\BibitemShut {NoStop}%
\bibitem [{\citenamefont {Karsch}\ \emph {et~al.}(2006)\citenamefont {Karsch},
  \citenamefont {Kharzeev},\ and\ \citenamefont {Satz}}]{Karsch:2005nk}%
  \BibitemOpen
  \bibfield  {author} {\bibinfo {author} {\bibfnamefont {F.}~\bibnamefont
  {Karsch}}, \bibinfo {author} {\bibfnamefont {D.}~\bibnamefont {Kharzeev}}, \
  and\ \bibinfo {author} {\bibfnamefont {H.}~\bibnamefont {Satz}},\ }\bibfield
  {title} {\enquote {\bibinfo {title} {{Sequential charmonium dissociation}},}\
  }\href {\doibase 10.1016/j.physletb.2006.03.078} {\bibfield  {journal}
  {\bibinfo  {journal} {Phys. Lett.}\ }\textbf {\bibinfo {volume} {B637}},\
  \bibinfo {pages} {75--80} (\bibinfo {year} {2006})},\ \Eprint
  {http://arxiv.org/abs/hep-ph/0512239} {arXiv:hep-ph/0512239 [hep-ph]}
  \BibitemShut {NoStop}%
\bibitem [{\citenamefont {Laine}\ \emph
  {et~al.}(2007{\natexlab{a}})\citenamefont {Laine}, \citenamefont {Philipsen},
  \citenamefont {Romatschke},\ and\ \citenamefont {Tassler}}]{Laine:2006ns}%
  \BibitemOpen
  \bibfield  {author} {\bibinfo {author} {\bibfnamefont {M.}~\bibnamefont
  {Laine}}, \bibinfo {author} {\bibfnamefont {O.}~\bibnamefont {Philipsen}},
  \bibinfo {author} {\bibfnamefont {P.}~\bibnamefont {Romatschke}}, \ and\
  \bibinfo {author} {\bibfnamefont {M.}~\bibnamefont {Tassler}},\ }\bibfield
  {title} {\enquote {\bibinfo {title} {{Real-time static potential in hot
  QCD}},}\ }\href {\doibase 10.1088/1126-6708/2007/03/054} {\bibfield
  {journal} {\bibinfo  {journal} {JHEP}\ }\textbf {\bibinfo {volume} {03}},\
  \bibinfo {pages} {054} (\bibinfo {year} {2007}{\natexlab{a}})},\ \Eprint
  {http://arxiv.org/abs/hep-ph/0611300} {arXiv:hep-ph/0611300 [hep-ph]}
  \BibitemShut {NoStop}%
\bibitem [{\citenamefont {Beraudo}\ \emph {et~al.}(2008)\citenamefont
  {Beraudo}, \citenamefont {Blaizot},\ and\ \citenamefont
  {Ratti}}]{Beraudo:2007ky}%
  \BibitemOpen
  \bibfield  {author} {\bibinfo {author} {\bibfnamefont {A.}~\bibnamefont
  {Beraudo}}, \bibinfo {author} {\bibfnamefont {J.~P.}\ \bibnamefont
  {Blaizot}}, \ and\ \bibinfo {author} {\bibfnamefont {C.}~\bibnamefont
  {Ratti}},\ }\bibfield  {title} {\enquote {\bibinfo {title} {{Real and
  imaginary-time Q anti-Q correlators in a thermal medium}},}\ }\href {\doibase
  10.1016/j.nuclphysa.2008.03.001} {\bibfield  {journal} {\bibinfo  {journal}
  {Nucl. Phys.}\ }\textbf {\bibinfo {volume} {A806}},\ \bibinfo {pages}
  {312--338} (\bibinfo {year} {2008})},\ \Eprint
  {http://arxiv.org/abs/0712.4394} {arXiv:0712.4394 [nucl-th]} \BibitemShut
  {NoStop}%
\bibitem [{\citenamefont {Burnier}\ \emph {et~al.}(2008)\citenamefont
  {Burnier}, \citenamefont {Laine},\ and\ \citenamefont
  {Vepsalainen}}]{Burnier:2007qm}%
  \BibitemOpen
  \bibfield  {author} {\bibinfo {author} {\bibfnamefont {Y.}~\bibnamefont
  {Burnier}}, \bibinfo {author} {\bibfnamefont {M.}~\bibnamefont {Laine}}, \
  and\ \bibinfo {author} {\bibfnamefont {M.}~\bibnamefont {Vepsalainen}},\
  }\bibfield  {title} {\enquote {\bibinfo {title} {{Heavy quarkonium in any
  channel in resummed hot QCD}},}\ }\href {\doibase
  10.1088/1126-6708/2008/01/043} {\bibfield  {journal} {\bibinfo  {journal}
  {JHEP}\ }\textbf {\bibinfo {volume} {01}},\ \bibinfo {pages} {043} (\bibinfo
  {year} {2008})},\ \Eprint {http://arxiv.org/abs/0711.1743} {arXiv:0711.1743
  [hep-ph]} \BibitemShut {NoStop}%
\bibitem [{\citenamefont {Laine}\ \emph
  {et~al.}(2007{\natexlab{b}})\citenamefont {Laine}, \citenamefont
  {Philipsen},\ and\ \citenamefont {Tassler}}]{Laine:2007qy}%
  \BibitemOpen
  \bibfield  {author} {\bibinfo {author} {\bibfnamefont {M.}~\bibnamefont
  {Laine}}, \bibinfo {author} {\bibfnamefont {O.}~\bibnamefont {Philipsen}}, \
  and\ \bibinfo {author} {\bibfnamefont {M.}~\bibnamefont {Tassler}},\
  }\bibfield  {title} {\enquote {\bibinfo {title} {{Thermal imaginary part of a
  real-time static potential from classical lattice gauge theory
  simulations}},}\ }\href {\doibase 10.1088/1126-6708/2007/09/066} {\bibfield
  {journal} {\bibinfo  {journal} {JHEP}\ }\textbf {\bibinfo {volume} {09}},\
  \bibinfo {pages} {066} (\bibinfo {year} {2007}{\natexlab{b}})},\ \Eprint
  {http://arxiv.org/abs/0707.2458} {arXiv:0707.2458 [hep-lat]} \BibitemShut
  {NoStop}%
\bibitem [{\citenamefont {Rothkopf}\ \emph {et~al.}(2012)\citenamefont
  {Rothkopf}, \citenamefont {Hatsuda},\ and\ \citenamefont
  {Sasaki}}]{Rothkopf:2011db}%
  \BibitemOpen
  \bibfield  {author} {\bibinfo {author} {\bibfnamefont {Alexander}\
  \bibnamefont {Rothkopf}}, \bibinfo {author} {\bibfnamefont {Tetsuo}\
  \bibnamefont {Hatsuda}}, \ and\ \bibinfo {author} {\bibfnamefont {Shoichi}\
  \bibnamefont {Sasaki}},\ }\bibfield  {title} {\enquote {\bibinfo {title}
  {{Complex Heavy-Quark Potential at Finite Temperature from Lattice QCD}},}\
  }\href {\doibase 10.1103/PhysRevLett.108.162001} {\bibfield  {journal}
  {\bibinfo  {journal} {Phys. Rev. Lett.}\ }\textbf {\bibinfo {volume} {108}},\
  \bibinfo {pages} {162001} (\bibinfo {year} {2012})},\ \Eprint
  {http://arxiv.org/abs/1108.1579} {arXiv:1108.1579 [hep-lat]} \BibitemShut
  {NoStop}%
\bibitem [{\citenamefont {Brambilla}\ \emph {et~al.}(2008)\citenamefont
  {Brambilla}, \citenamefont {Ghiglieri}, \citenamefont {Vairo},\ and\
  \citenamefont {Petreczky}}]{Brambilla:2008cx}%
  \BibitemOpen
  \bibfield  {author} {\bibinfo {author} {\bibfnamefont {Nora}\ \bibnamefont
  {Brambilla}}, \bibinfo {author} {\bibfnamefont {Jacopo}\ \bibnamefont
  {Ghiglieri}}, \bibinfo {author} {\bibfnamefont {Antonio}\ \bibnamefont
  {Vairo}}, \ and\ \bibinfo {author} {\bibfnamefont {Peter}\ \bibnamefont
  {Petreczky}},\ }\bibfield  {title} {\enquote {\bibinfo {title} {{Static
  quark-antiquark pairs at finite temperature}},}\ }\href {\doibase
  10.1103/PhysRevD.78.014017} {\bibfield  {journal} {\bibinfo  {journal} {Phys.
  Rev.}\ }\textbf {\bibinfo {volume} {D78}},\ \bibinfo {pages} {014017}
  (\bibinfo {year} {2008})},\ \Eprint {http://arxiv.org/abs/0804.0993}
  {arXiv:0804.0993 [hep-ph]} \BibitemShut {NoStop}%
\bibitem [{\citenamefont {Andronic}\ \emph {et~al.}(2016)\citenamefont
  {Andronic} \emph {et~al.}}]{Andronic:2015wma}%
  \BibitemOpen
  \bibfield  {author} {\bibinfo {author} {\bibfnamefont {A.}~\bibnamefont
  {Andronic}} \emph {et~al.},\ }\bibfield  {title} {\enquote {\bibinfo {title}
  {{Heavy-flavour and quarkonium production in the LHC era: from
  proton–proton to heavy-ion collisions}},}\ }\href {\doibase
  10.1140/epjc/s10052-015-3819-5} {\bibfield  {journal} {\bibinfo  {journal}
  {Eur. Phys. J.}\ }\textbf {\bibinfo {volume} {C76}},\ \bibinfo {pages} {107}
  (\bibinfo {year} {2016})},\ \Eprint {http://arxiv.org/abs/1506.03981}
  {arXiv:1506.03981 [nucl-ex]} \BibitemShut {NoStop}%
\bibitem [{\citenamefont {Aarts}\ \emph {et~al.}(2017)\citenamefont {Aarts}
  \emph {et~al.}}]{Aarts:2016hap}%
  \BibitemOpen
  \bibfield  {author} {\bibinfo {author} {\bibfnamefont {G.}~\bibnamefont
  {Aarts}} \emph {et~al.},\ }\bibfield  {title} {\enquote {\bibinfo {title}
  {{Heavy-flavor production and medium properties in high-energy nuclear
  collisions - What next?}}}\ }\href {\doibase 10.1140/epja/i2017-12282-9}
  {\bibfield  {journal} {\bibinfo  {journal} {Eur. Phys. J.}\ }\textbf
  {\bibinfo {volume} {A53}},\ \bibinfo {pages} {93} (\bibinfo {year} {2017})},\
  \Eprint {http://arxiv.org/abs/1612.08032} {arXiv:1612.08032 [nucl-th]}
  \BibitemShut {NoStop}%
\bibitem [{\citenamefont {Rey}\ \emph {et~al.}(1998)\citenamefont {Rey},
  \citenamefont {Theisen},\ and\ \citenamefont {Yee}}]{Rey:1998bq}%
  \BibitemOpen
  \bibfield  {author} {\bibinfo {author} {\bibfnamefont {Soo-Jong}\
  \bibnamefont {Rey}}, \bibinfo {author} {\bibfnamefont {Stefan}\ \bibnamefont
  {Theisen}}, \ and\ \bibinfo {author} {\bibfnamefont {Jung-Tay}\ \bibnamefont
  {Yee}},\ }\bibfield  {title} {\enquote {\bibinfo {title} {{Wilson-Polyakov
  loop at finite temperature in large N gauge theory and anti-de Sitter
  supergravity}},}\ }\href {\doibase 10.1016/S0550-3213(98)00471-4} {\bibfield
  {journal} {\bibinfo  {journal} {Nucl. Phys.}\ }\textbf {\bibinfo {volume}
  {B527}},\ \bibinfo {pages} {171--186} (\bibinfo {year} {1998})},\ \Eprint
  {http://arxiv.org/abs/hep-th/9803135} {arXiv:hep-th/9803135 [hep-th]}
  \BibitemShut {NoStop}%
\bibitem [{\citenamefont {Brandhuber}\ \emph {et~al.}(1998)\citenamefont
  {Brandhuber}, \citenamefont {Itzhaki}, \citenamefont {Sonnenschein},\ and\
  \citenamefont {Yankielowicz}}]{Brandhuber:1998bs}%
  \BibitemOpen
  \bibfield  {author} {\bibinfo {author} {\bibfnamefont {A.}~\bibnamefont
  {Brandhuber}}, \bibinfo {author} {\bibfnamefont {N.}~\bibnamefont {Itzhaki}},
  \bibinfo {author} {\bibfnamefont {J.}~\bibnamefont {Sonnenschein}}, \ and\
  \bibinfo {author} {\bibfnamefont {S.}~\bibnamefont {Yankielowicz}},\
  }\bibfield  {title} {\enquote {\bibinfo {title} {{Wilson loops in the large N
  limit at finite temperature}},}\ }\href {\doibase
  10.1016/S0370-2693(98)00730-8} {\bibfield  {journal} {\bibinfo  {journal}
  {Phys. Lett.}\ }\textbf {\bibinfo {volume} {B434}},\ \bibinfo {pages}
  {36--40} (\bibinfo {year} {1998})},\ \Eprint
  {http://arxiv.org/abs/hep-th/9803137} {arXiv:hep-th/9803137 [hep-th]}
  \BibitemShut {NoStop}%
\bibitem [{\citenamefont {Albacete}\ \emph {et~al.}(2008)\citenamefont
  {Albacete}, \citenamefont {Kovchegov},\ and\ \citenamefont
  {Taliotis}}]{Albacete:2008dz}%
  \BibitemOpen
  \bibfield  {author} {\bibinfo {author} {\bibfnamefont {Javier~L.}\
  \bibnamefont {Albacete}}, \bibinfo {author} {\bibfnamefont {Yuri~V.}\
  \bibnamefont {Kovchegov}}, \ and\ \bibinfo {author} {\bibfnamefont
  {Anastasios}\ \bibnamefont {Taliotis}},\ }\bibfield  {title} {\enquote
  {\bibinfo {title} {{Heavy Quark Potential at Finite Temperature in AdS/CFT
  Revisited}},}\ }\href {\doibase 10.1103/PhysRevD.78.115007} {\bibfield
  {journal} {\bibinfo  {journal} {Phys. Rev.}\ }\textbf {\bibinfo {volume}
  {D78}},\ \bibinfo {pages} {115007} (\bibinfo {year} {2008})},\ \Eprint
  {http://arxiv.org/abs/0807.4747} {arXiv:0807.4747 [hep-th]} \BibitemShut
  {NoStop}%
\bibitem [{\citenamefont {Noronha}\ and\ \citenamefont
  {Dumitru}(2009)}]{Noronha:2009da}%
  \BibitemOpen
  \bibfield  {author} {\bibinfo {author} {\bibfnamefont {Jorge}\ \bibnamefont
  {Noronha}}\ and\ \bibinfo {author} {\bibfnamefont {Adrian}\ \bibnamefont
  {Dumitru}},\ }\bibfield  {title} {\enquote {\bibinfo {title} {{Thermal Width
  of the $\Upsilon$ at Large t' Hooft Coupling}},}\ }\href {\doibase
  10.1103/PhysRevLett.103.152304} {\bibfield  {journal} {\bibinfo  {journal}
  {Phys. Rev. Lett.}\ }\textbf {\bibinfo {volume} {103}},\ \bibinfo {pages}
  {152304} (\bibinfo {year} {2009})},\ \Eprint {http://arxiv.org/abs/0907.3062}
  {arXiv:0907.3062 [hep-ph]} \BibitemShut {NoStop}%
\bibitem [{\citenamefont {Hayata}\ \emph {et~al.}(2013)\citenamefont {Hayata},
  \citenamefont {Nawa},\ and\ \citenamefont {Hatsuda}}]{Hayata:2012rw}%
  \BibitemOpen
  \bibfield  {author} {\bibinfo {author} {\bibfnamefont {Tomoya}\ \bibnamefont
  {Hayata}}, \bibinfo {author} {\bibfnamefont {Kanabu}\ \bibnamefont {Nawa}}, \
  and\ \bibinfo {author} {\bibfnamefont {Tetsuo}\ \bibnamefont {Hatsuda}},\
  }\bibfield  {title} {\enquote {\bibinfo {title} {{Time-dependent heavy-quark
  potential at finite temperature from gauge-gravity duality}},}\ }\href
  {\doibase 10.1103/PhysRevD.87.101901} {\bibfield  {journal} {\bibinfo
  {journal} {Phys. Rev.}\ }\textbf {\bibinfo {volume} {D87}},\ \bibinfo {pages}
  {101901} (\bibinfo {year} {2013})},\ \Eprint {http://arxiv.org/abs/1211.4942}
  {arXiv:1211.4942 [hep-ph]} \BibitemShut {NoStop}%
\bibitem [{\citenamefont {Bitaghsir~Fadafan}\ and\ \citenamefont
  {Tabatabaei}(2016)}]{Fadafan:2015kma}%
  \BibitemOpen
  \bibfield  {author} {\bibinfo {author} {\bibfnamefont {Kazem}\ \bibnamefont
  {Bitaghsir~Fadafan}}\ and\ \bibinfo {author} {\bibfnamefont {Seyed~Kamal}\
  \bibnamefont {Tabatabaei}},\ }\bibfield  {title} {\enquote {\bibinfo {title}
  {{The Imaginary Potential and Thermal Width of Moving Quarkonium from
  Holography}},}\ }\href {\doibase 10.1088/0954-3899/43/9/095001} {\bibfield
  {journal} {\bibinfo  {journal} {J. Phys.}\ }\textbf {\bibinfo {volume}
  {G43}},\ \bibinfo {pages} {095001} (\bibinfo {year} {2016})},\ \Eprint
  {http://arxiv.org/abs/1501.00439} {arXiv:1501.00439 [hep-th]} \BibitemShut
  {NoStop}%
\bibitem [{\citenamefont {Liu}\ \emph {et~al.}(2007)\citenamefont {Liu},
  \citenamefont {Rajagopal},\ and\ \citenamefont {Wiedemann}}]{Liu:2006nn}%
  \BibitemOpen
  \bibfield  {author} {\bibinfo {author} {\bibfnamefont {Hong}\ \bibnamefont
  {Liu}}, \bibinfo {author} {\bibfnamefont {Krishna}\ \bibnamefont
  {Rajagopal}}, \ and\ \bibinfo {author} {\bibfnamefont {Urs~Achim}\
  \bibnamefont {Wiedemann}},\ }\bibfield  {title} {\enquote {\bibinfo {title}
  {{An AdS/CFT Calculation of Screening in a Hot Wind}},}\ }\href {\doibase
  10.1103/PhysRevLett.98.182301} {\bibfield  {journal} {\bibinfo  {journal}
  {Phys. Rev. Lett.}\ }\textbf {\bibinfo {volume} {98}},\ \bibinfo {pages}
  {182301} (\bibinfo {year} {2007})},\ \Eprint
  {http://arxiv.org/abs/hep-ph/0607062} {arXiv:hep-ph/0607062 [hep-ph]}
  \BibitemShut {NoStop}%
\bibitem [{\citenamefont {Finazzo}\ and\ \citenamefont
  {Noronha}(2015)}]{Finazzo:2014rca}%
  \BibitemOpen
  \bibfield  {author} {\bibinfo {author} {\bibfnamefont {Stefano~I.}\
  \bibnamefont {Finazzo}}\ and\ \bibinfo {author} {\bibfnamefont {Jorge}\
  \bibnamefont {Noronha}},\ }\bibfield  {title} {\enquote {\bibinfo {title}
  {{Thermal suppression of moving heavy quark pairs in a strongly coupled
  plasma}},}\ }\href {\doibase 10.1007/JHEP01(2015)051} {\bibfield  {journal}
  {\bibinfo  {journal} {JHEP}\ }\textbf {\bibinfo {volume} {01}},\ \bibinfo
  {pages} {051} (\bibinfo {year} {2015})},\ \Eprint
  {http://arxiv.org/abs/1406.2683} {arXiv:1406.2683 [hep-th]} \BibitemShut
  {NoStop}%
\bibitem [{\citenamefont {Braga}\ and\ \citenamefont
  {Ferreira}(2016)}]{Braga:2016oem}%
  \BibitemOpen
  \bibfield  {author} {\bibinfo {author} {\bibfnamefont {Nelson R.~F.}\
  \bibnamefont {Braga}}\ and\ \bibinfo {author} {\bibfnamefont {Luiz~F.}\
  \bibnamefont {Ferreira}},\ }\bibfield  {title} {\enquote {\bibinfo {title}
  {{Thermal width of heavy quarkonia from an AdS/QCD model}},}\ }\href
  {\doibase 10.1103/PhysRevD.94.094019} {\bibfield  {journal} {\bibinfo
  {journal} {Phys. Rev.}\ }\textbf {\bibinfo {volume} {D94}},\ \bibinfo {pages}
  {094019} (\bibinfo {year} {2016})},\ \Eprint
  {http://arxiv.org/abs/1606.09535} {arXiv:1606.09535 [hep-th]} \BibitemShut
  {NoStop}%
\bibitem [{\citenamefont {Patra}\ \emph {et~al.}(2015)\citenamefont {Patra},
  \citenamefont {Khanchandani},\ and\ \citenamefont {Thakur}}]{Patra:2015qoa}%
  \BibitemOpen
  \bibfield  {author} {\bibinfo {author} {\bibfnamefont {Binoy~Krishna}\
  \bibnamefont {Patra}}, \bibinfo {author} {\bibfnamefont {Himanshu}\
  \bibnamefont {Khanchandani}}, \ and\ \bibinfo {author} {\bibfnamefont {Lata}\
  \bibnamefont {Thakur}},\ }\bibfield  {title} {\enquote {\bibinfo {title}
  {{Velocity-induced Heavy Quarkonium Dissociation using the gauge-gravity
  correspondence}},}\ }\href {\doibase 10.1103/PhysRevD.92.085034} {\bibfield
  {journal} {\bibinfo  {journal} {Phys. Rev.}\ }\textbf {\bibinfo {volume}
  {D92}},\ \bibinfo {pages} {085034} (\bibinfo {year} {2015})},\ \Eprint
  {http://arxiv.org/abs/1504.05396} {arXiv:1504.05396 [hep-th]} \BibitemShut
  {NoStop}%
\bibitem [{\citenamefont {Ali-Akbari}\ \emph {et~al.}(2014)\citenamefont
  {Ali-Akbari}, \citenamefont {Giataganas},\ and\ \citenamefont
  {Rezaei}}]{Ali-Akbari:2014vpa}%
  \BibitemOpen
  \bibfield  {author} {\bibinfo {author} {\bibfnamefont {M.}~\bibnamefont
  {Ali-Akbari}}, \bibinfo {author} {\bibfnamefont {D.}~\bibnamefont
  {Giataganas}}, \ and\ \bibinfo {author} {\bibfnamefont {Z.}~\bibnamefont
  {Rezaei}},\ }\bibfield  {title} {\enquote {\bibinfo {title} {{Imaginary
  potential of heavy quarkonia moving in strongly coupled plasma}},}\ }\href
  {\doibase 10.1103/PhysRevD.90.086001} {\bibfield  {journal} {\bibinfo
  {journal} {Phys. Rev.}\ }\textbf {\bibinfo {volume} {D90}},\ \bibinfo {pages}
  {086001} (\bibinfo {year} {2014})},\ \Eprint {http://arxiv.org/abs/1406.1994}
  {arXiv:1406.1994 [hep-th]} \BibitemShut {NoStop}%
\bibitem [{\citenamefont {Escobedo}\ \emph {et~al.}(2013)\citenamefont
  {Escobedo}, \citenamefont {Giannuzzi}, \citenamefont {Mannarelli},\ and\
  \citenamefont {Soto}}]{Escobedo:2013tca}%
  \BibitemOpen
  \bibfield  {author} {\bibinfo {author} {\bibfnamefont {Miguel~Angel}\
  \bibnamefont {Escobedo}}, \bibinfo {author} {\bibfnamefont {Floriana}\
  \bibnamefont {Giannuzzi}}, \bibinfo {author} {\bibfnamefont {Massimo}\
  \bibnamefont {Mannarelli}}, \ and\ \bibinfo {author} {\bibfnamefont {Joan}\
  \bibnamefont {Soto}},\ }\bibfield  {title} {\enquote {\bibinfo {title}
  {{Heavy Quarkonium moving in a Quark-Gluon Plasma}},}\ }\href {\doibase
  10.1103/PhysRevD.87.114005} {\bibfield  {journal} {\bibinfo  {journal} {Phys.
  Rev.}\ }\textbf {\bibinfo {volume} {D87}},\ \bibinfo {pages} {114005}
  (\bibinfo {year} {2013})},\ \Eprint {http://arxiv.org/abs/1304.4087}
  {arXiv:1304.4087 [hep-ph]} \BibitemShut {NoStop}%
\bibitem [{\citenamefont {Krouppa}\ \emph {et~al.}(2015)\citenamefont
  {Krouppa}, \citenamefont {Ryblewski},\ and\ \citenamefont
  {Strickland}}]{Krouppa:2015yoa}%
  \BibitemOpen
  \bibfield  {author} {\bibinfo {author} {\bibfnamefont {Brandon}\ \bibnamefont
  {Krouppa}}, \bibinfo {author} {\bibfnamefont {Radoslaw}\ \bibnamefont
  {Ryblewski}}, \ and\ \bibinfo {author} {\bibfnamefont {Michael}\ \bibnamefont
  {Strickland}},\ }\bibfield  {title} {\enquote {\bibinfo {title} {{Bottomonia
  suppression in 2.76 TeV Pb-Pb collisions}},}\ }\href {\doibase
  10.1103/PhysRevC.92.061901} {\bibfield  {journal} {\bibinfo  {journal} {Phys.
  Rev.}\ }\textbf {\bibinfo {volume} {C92}},\ \bibinfo {pages} {061901}
  (\bibinfo {year} {2015})},\ \Eprint {http://arxiv.org/abs/1507.03951}
  {arXiv:1507.03951 [hep-ph]} \BibitemShut {NoStop}%
\bibitem [{\citenamefont {Kraft}(2013)}]{Kraft2013}%
  \BibitemOpen
  \bibfield  {author} {\bibinfo {author} {\bibfnamefont {D}~\bibnamefont
  {Kraft}},\ }\emph {\bibinfo {title} {Stochastic Variational Approaches to
  Non-Hermitian Quantum-Mechanical Problems}},\ \href@noop {} {Master's
  thesis},\ \bibinfo  {school} {Graz University} (\bibinfo {year}
  {2013})\BibitemShut {NoStop}%
\bibitem [{\citenamefont {Strickland}\ and\ \citenamefont
  {Bazow}(2012)}]{Strickland:2011aa}%
  \BibitemOpen
  \bibfield  {author} {\bibinfo {author} {\bibfnamefont {Michael}\ \bibnamefont
  {Strickland}}\ and\ \bibinfo {author} {\bibfnamefont {Dennis}\ \bibnamefont
  {Bazow}},\ }\bibfield  {title} {\enquote {\bibinfo {title} {{Thermal
  Bottomonium Suppression at RHIC and LHC}},}\ }\href {\doibase
  10.1016/j.nuclphysa.2012.02.003} {\bibfield  {journal} {\bibinfo  {journal}
  {Nucl. Phys.}\ }\textbf {\bibinfo {volume} {A879}},\ \bibinfo {pages}
  {25--58} (\bibinfo {year} {2012})},\ \Eprint {http://arxiv.org/abs/1112.2761}
  {arXiv:1112.2761 [nucl-th]} \BibitemShut {NoStop}%
\bibitem [{\citenamefont {Dumitru}\ \emph {et~al.}(2009)\citenamefont
  {Dumitru}, \citenamefont {Guo},\ and\ \citenamefont
  {Strickland}}]{Dumitru:2009fy}%
  \BibitemOpen
  \bibfield  {author} {\bibinfo {author} {\bibfnamefont {Adrian}\ \bibnamefont
  {Dumitru}}, \bibinfo {author} {\bibfnamefont {Yun}\ \bibnamefont {Guo}}, \
  and\ \bibinfo {author} {\bibfnamefont {Michael}\ \bibnamefont {Strickland}},\
  }\bibfield  {title} {\enquote {\bibinfo {title} {{The Imaginary part of the
  static gluon propagator in an anisotropic (viscous) QCD plasma}},}\ }\href
  {\doibase 10.1103/PhysRevD.79.114003} {\bibfield  {journal} {\bibinfo
  {journal} {Phys. Rev.}\ }\textbf {\bibinfo {volume} {D79}},\ \bibinfo {pages}
  {114003} (\bibinfo {year} {2009})},\ \Eprint {http://arxiv.org/abs/0903.4703}
  {arXiv:0903.4703 [hep-ph]} \BibitemShut {NoStop}%
\bibitem [{\citenamefont {Kaczmarek}\ \emph {et~al.}(2004)\citenamefont
  {Kaczmarek}, \citenamefont {Karsch}, \citenamefont {Zantow},\ and\
  \citenamefont {Petreczky}}]{Kaczmarek:2004gv}%
  \BibitemOpen
  \bibfield  {author} {\bibinfo {author} {\bibfnamefont {O.}~\bibnamefont
  {Kaczmarek}}, \bibinfo {author} {\bibfnamefont {F.}~\bibnamefont {Karsch}},
  \bibinfo {author} {\bibfnamefont {F.}~\bibnamefont {Zantow}}, \ and\ \bibinfo
  {author} {\bibfnamefont {P.}~\bibnamefont {Petreczky}},\ }\bibfield  {title}
  {\enquote {\bibinfo {title} {{Static quark anti-quark free energy and the
  running coupling at finite temperature}},}\ }\href {\doibase
  10.1103/PhysRevD.70.074505, 10.1103/PhysRevD.72.059903} {\bibfield  {journal}
  {\bibinfo  {journal} {Phys. Rev.}\ }\textbf {\bibinfo {volume} {D70}},\
  \bibinfo {pages} {074505} (\bibinfo {year} {2004})},\ \bibinfo {note}
  {[Erratum: Phys. Rev.D72,059903(2005)]},\ \Eprint
  {http://arxiv.org/abs/hep-lat/0406036} {arXiv:hep-lat/0406036 [hep-lat]}
  \BibitemShut {NoStop}%
\bibitem [{\citenamefont {Margotta}\ \emph {et~al.}(2011)\citenamefont
  {Margotta}, \citenamefont {McCarty}, \citenamefont {McGahan}, \citenamefont
  {Strickland},\ and\ \citenamefont {Yager-Elorriaga}}]{Margotta:2011ta}%
  \BibitemOpen
  \bibfield  {author} {\bibinfo {author} {\bibfnamefont {Matthew}\ \bibnamefont
  {Margotta}}, \bibinfo {author} {\bibfnamefont {Kyle}\ \bibnamefont
  {McCarty}}, \bibinfo {author} {\bibfnamefont {Christina}\ \bibnamefont
  {McGahan}}, \bibinfo {author} {\bibfnamefont {Michael}\ \bibnamefont
  {Strickland}}, \ and\ \bibinfo {author} {\bibfnamefont {David}\ \bibnamefont
  {Yager-Elorriaga}},\ }\bibfield  {title} {\enquote {\bibinfo {title}
  {{Quarkonium states in a complex-valued potential}},}\ }\href {\doibase
  10.1103/PhysRevD.84.069902, 10.1103/PhysRevD.83.105019} {\bibfield  {journal}
  {\bibinfo  {journal} {Phys. Rev.}\ }\textbf {\bibinfo {volume} {D83}},\
  \bibinfo {pages} {105019} (\bibinfo {year} {2011})},\ \bibinfo {note}
  {[Erratum: Phys. Rev.D84,069902(2011)]},\ \Eprint
  {http://arxiv.org/abs/1101.4651} {arXiv:1101.4651 [hep-ph]} \BibitemShut
  {NoStop}%
\bibitem [{\citenamefont {Maldacena}(1998)}]{Maldacena:1998im}%
  \BibitemOpen
  \bibfield  {author} {\bibinfo {author} {\bibfnamefont {Juan~Martin}\
  \bibnamefont {Maldacena}},\ }\bibfield  {title} {\enquote {\bibinfo {title}
  {{Wilson loops in large N field theories}},}\ }\href {\doibase
  10.1103/PhysRevLett.80.4859} {\bibfield  {journal} {\bibinfo  {journal}
  {Phys. Rev. Lett.}\ }\textbf {\bibinfo {volume} {80}},\ \bibinfo {pages}
  {4859--4862} (\bibinfo {year} {1998})},\ \Eprint
  {http://arxiv.org/abs/hep-th/9803002} {arXiv:hep-th/9803002 [hep-th]}
  \BibitemShut {NoStop}%
\bibitem [{\citenamefont {Sudiarta}\ and\ \citenamefont
  {Geldart}(2007)}]{1751-8121-40-8-013}%
  \BibitemOpen
  \bibfield  {author} {\bibinfo {author} {\bibfnamefont {I~Wayan}\ \bibnamefont
  {Sudiarta}}\ and\ \bibinfo {author} {\bibfnamefont {D~J~Wallace}\
  \bibnamefont {Geldart}},\ }\bibfield  {title} {\enquote {\bibinfo {title}
  {Solving the schrödinger equation using the finite difference time domain
  method},}\ }\href {http://stacks.iop.org/1751-8121/40/i=8/a=013} {\bibfield
  {journal} {\bibinfo  {journal} {Journal of Physics A: Mathematical and
  Theoretical}\ }\textbf {\bibinfo {volume} {40}},\ \bibinfo {pages} {1885}
  (\bibinfo {year} {2007})}\BibitemShut {NoStop}%
\bibitem [{\citenamefont {Press}\ \emph {et~al.}(2007)\citenamefont {Press},
  \citenamefont {Teukolsky}, \citenamefont {Vetterling},\ and\ \citenamefont
  {Flannery}}]{Press:2007:NRE:1403886}%
  \BibitemOpen
  \bibfield  {author} {\bibinfo {author} {\bibfnamefont {William~H.}\
  \bibnamefont {Press}}, \bibinfo {author} {\bibfnamefont {Saul~A.}\
  \bibnamefont {Teukolsky}}, \bibinfo {author} {\bibfnamefont {William~T.}\
  \bibnamefont {Vetterling}}, \ and\ \bibinfo {author} {\bibfnamefont
  {Brian~P.}\ \bibnamefont {Flannery}},\ }\href@noop {} {\emph {\bibinfo
  {title} {Numerical Recipes 3rd Edition: The Art of Scientific Computing}}},\
  \bibinfo {edition} {3rd}\ ed.\ (\bibinfo  {publisher} {Cambridge University
  Press},\ \bibinfo {address} {New York, NY, USA},\ \bibinfo {year} {2007})\
  pp.\ \bibinfo {pages} {1043--1046}\BibitemShut {NoStop}%
\bibitem [{\citenamefont {Khachatryan}\ \emph {et~al.}(2017)\citenamefont
  {Khachatryan} \emph {et~al.}}]{Khachatryan:2016xxp}%
  \BibitemOpen
  \bibfield  {author} {\bibinfo {author} {\bibfnamefont {Vardan}\ \bibnamefont
  {Khachatryan}} \emph {et~al.} (\bibinfo {collaboration} {CMS}),\ }\bibfield
  {title} {\enquote {\bibinfo {title} {{Suppression of $\Upsilon(1S),
  \Upsilon(2S)$ and $\Upsilon(3S)$ production in PbPb collisions at
  $\sqrt{s_{\rm NN}}$ = 2.76 TeV}},}\ }\href {\doibase
  10.1016/j.physletb.2017.04.031} {\bibfield  {journal} {\bibinfo  {journal}
  {Phys. Lett.}\ }\textbf {\bibinfo {volume} {B770}},\ \bibinfo {pages}
  {357--379} (\bibinfo {year} {2017})},\ \Eprint
  {http://arxiv.org/abs/1611.01510} {arXiv:1611.01510 [nucl-ex]} \BibitemShut
  {NoStop}%
\bibitem [{\citenamefont {Miller}\ \emph {et~al.}(2007)\citenamefont {Miller},
  \citenamefont {Reygers}, \citenamefont {Sanders},\ and\ \citenamefont
  {Steinberg}}]{Miller:2007ri}%
  \BibitemOpen
  \bibfield  {author} {\bibinfo {author} {\bibfnamefont {Michael~L.}\
  \bibnamefont {Miller}}, \bibinfo {author} {\bibfnamefont {Klaus}\
  \bibnamefont {Reygers}}, \bibinfo {author} {\bibfnamefont {Stephen~J.}\
  \bibnamefont {Sanders}}, \ and\ \bibinfo {author} {\bibfnamefont {Peter}\
  \bibnamefont {Steinberg}},\ }\bibfield  {title} {\enquote {\bibinfo {title}
  {{Glauber modeling in high energy nuclear collisions}},}\ }\href {\doibase
  10.1146/annurev.nucl.57.090506.123020} {\bibfield  {journal} {\bibinfo
  {journal} {Ann. Rev. Nucl. Part. Sci.}\ }\textbf {\bibinfo {volume} {57}},\
  \bibinfo {pages} {205--243} (\bibinfo {year} {2007})},\ \Eprint
  {http://arxiv.org/abs/nucl-ex/0701025} {arXiv:nucl-ex/0701025 [nucl-ex]}
  \BibitemShut {NoStop}%
\bibitem [{\citenamefont {Gubser}(2008)}]{Gubser:2006nz}%
  \BibitemOpen
  \bibfield  {author} {\bibinfo {author} {\bibfnamefont {Steven~S.}\
  \bibnamefont {Gubser}},\ }\bibfield  {title} {\enquote {\bibinfo {title}
  {{Momentum fluctuations of heavy quarks in the gauge-string duality}},}\
  }\href {\doibase 10.1016/j.nuclphysb.2007.09.017} {\bibfield  {journal}
  {\bibinfo  {journal} {Nucl. Phys.}\ }\textbf {\bibinfo {volume} {B790}},\
  \bibinfo {pages} {175--199} (\bibinfo {year} {2008})},\ \Eprint
  {http://arxiv.org/abs/hep-th/0612143} {arXiv:hep-th/0612143 [hep-th]}
  \BibitemShut {NoStop}%
\bibitem [{\citenamefont {Messiah}(2014)}]{Messiah2014}%
  \BibitemOpen
  \bibfield  {author} {\bibinfo {author} {\bibfnamefont {Albert}\ \bibnamefont
  {Messiah}},\ }\href@noop {} {\emph {\bibinfo {title} {Quantum Mechanics}}}\
  (\bibinfo  {publisher} {Dover Publications},\ \bibinfo {year}
  {2014})\BibitemShut {NoStop}%
\end{thebibliography}
\end{document}